\documentclass{article}
\usepackage{amsmath}
\usepackage{amssymb}
\usepackage{graphicx}
\usepackage{cite}
\usepackage{booktabs}
\usepackage{caption}
\usepackage{array}
\title{Recent advances in cosmological singularities}
\author{Oem Trivedi \footnote{oem.t@ahduni.edu.in}}
\date{%
	School of Arts and Sciences, Ahmedabad Univeristy, Ahmedabad 380009, India 
	\today
}

\begin{document}
	
	\maketitle

	\begin{abstract}
		The discovery of universe's late-time acceleration and dark energy has overseen a great deal of research into cosmological singularities and in this brief review, we discuss all the prominent developments in this field for the best part of the last 2 decades. We discuss the fundamentals of space-time singularities after which we discuss about all the different forms of cosmological singularities which have been discovered in recent times in detail. We then talk about methods and techniques to avoid or moderate these singularities in various theories and discuss how these singularities can occur in non-conventional cosmologies too. We then discuss a useful dynamical systems approach to deal with these singularities and finish up with some outlooks for the field. We hope that this work serves as a good resource to anyone who wants to update themselves with the developments in this very exciting area.   
	\end{abstract}

\section{Introduction}

Observations of late time acceleration of the Universe came as a huge surprise to the cosmological community \cite{SupernovaSearchTeam:1998fmf} and ever since then a lot of work has been done in order to explain this expansion. The cosmological expansion problem has been addressed from multiple facets till now, which include the standard approaches of the Cosmological constant \cite{Weinberg:1988cp,Lombriser:2019jia,Padmanabhan:2002ji} alongside more exotic scenarios like Modified gravity theories\cite{Capozziello:2011et,Nojiri:2010wj,Nojiri:2017ncd}, scalar field driven late-time cosmic acceleration scenarios \cite{Zlatev:1998tr,Tsujikawa:2013fta,Faraoni:2000wk,Gasperini:2001pc,Capozziello:2003tk,Capozziello:2002rd,Odintsov:2023weg}. Several approaches to quantum gravity have also weighed in on the cosmic-acceleration puzzle, ranging from the Braneworld cosmology of string theory to the likes of loop quantum cosmology and asymptotically safe cosmology \cite{Sahni:2002dx,Sami:2004xk,Tretyakov:2005en,Chen:2008ca,Fu:2008gh,Bonanno:2001hi,Bonanno:2001xi,Bentivegna:2003rr,Reuter:2005kb,Bonanno:2007wg,Weinberg:2009wa}. This, however, has also sprung up some discrepancies which seem to be pointing towards the limits of our current understanding of the universe, most famous of which is arguably the Hubble tension which refers to the disagreements between the values of the Hubble constant measured from.  Detailed Cosmic Microwave Background Radiation(CMB) maps, combined with Baryon Acoustic Oscillations data, and those from the SNeIa data \cite{Planck:2018vyg,riess2019large,riess2021comprehensive}. Hence, the current epoch of the universe has certainly provided us with a wide range of questions and looks set to become an avenue where advanced gravitational physics will lead the way towards better understanding of cosmology.
\\
\\
There has also been an expansive literature in recent times which has been devoted to the study of various types of singularities that could occur during the current and far future of the Universe, with the observation of late-time acceleration having given a significant boost to such works \cite{Nojiri:2004ip,Nojiri:2005sr,Nojiri:2005sx,Bamba:2008ut,trivedi2022finite,trivedi2022type,odintsov2015singular,odintsov2016singular,oikonomou2015singular,nojiri2015singular,Odintsov:2022eqm,de2023finite,Odintsov:2023qfj,Odintsov:2022qnn,Odintsov:2021yva,Nojiri:2022xdo,Brevik:2021wzs,Odintsov:2022unp,Giovannetti:2020nte,Barca:2021qdn,sym15091701,Choudhury:2011jt,Choudhury:2013zna}.Even
the term singularity comprises many different definitions and with regards to cosmological cases, until the
end of the 20th century, the only popular possibilites of singularty formation were the initial Big
Bang singularity and, in the case of spatially closed cosmological models, the final Big Crunch
singularity. The definition of a singular point in cosmology was given by Hawking
and Penrose , and most of the theorems proven by them make use of the null energy condition and also of
the facts that at a singular point of the spacetime, geodesics incompleteness occurs and also the curvature scalars
diverge. Although in modified gravity the null energy condition may be different in general in comparison to the
Einstein-Hilbert case (see for example), it is generally accepted that the geodesic incompleteness and also that
the divergence of the curvature invariants, strongly indicate the presence of a crushing singularity. The singularities
in cosmology vary in their effects, and a complete classification of these was performed in . While one can treat singularities as points at which a cosmological theory somewhat fails, one might also consider them as windows to new physics and thus have a different kind of appealing interest with it. Particularly finite-time singularities (those which happen in a finite time) could be viewed as
either flaws the classical theory, or alternatively as a doorway towards a quantum description of general relativity. This is due to the fact that these cannot be dressed in a similar way to the spacelike
singularities of black holes for instance, and so one is left to ponder about the accuracy of the predictions of classical gravitational theories. Hence studying singularities in cosmological contexts and how they could be (possibly) removed, provides a way towards a deeper understanding of relating quantum descriptions of cosmology with its classical ones.
\\
\\
These cosmological singularities which have been discussed in recent times can be classified broadly into two types ; strong and weak (such a classification was initially put forward by \cite{ellis1977singular}). Strong singularities are those singularities which can distort finite objects and can mark either the beginning or the end of the universe, with the big bang being the one for the start of the universe and the so called "big rip" signalling the end of the universe. Weak singularities, as the name might suggest, are those which do not have such far reaching implications and do not distort finite objects in the same sense as their strong counterparts. We can discuss these various singularities in more detail as follows, in accordance to the classification provided in \cite{Bamba:2008ut,capozziello2009classifying} : \begin{itemize}
	\setlength\itemsep{1em}
	\item Type -1 ("Grand Bang/ Grand rip") : In this scale the scale factor becomes null (Bang) or diverges (rip ) for $w=-1$ \cite{fernandez2014grand}
	\item Type 0 ("Big Bang") : In this case, the scale factor becomes null for  $w \neq  -1$  
	\item Type I ("Big Rip") : In this case, the scale factor , effective energy density and effective pressure density diverges for $w \neq -1$. This is a scenario of universal death, wherein everything which resides in the universe is progressively torn apart \cite{Caldwell:2003vq}.
	\item Type II ("Sudden/Quiescent singularity") : In this case, the pressure density diverges and so does the derivatives of the scalar factor from the second derivative onwards \cite{Barrow:2004xh}. The weak and strong energy conditions hold for this singularity. Also known as quiescent singularities, but this name originally appeared in contexts related to non-oscillatory singularities \cite{Andersson:2000cv}. A special case of this is the big brake singularity \cite{Gorini:2003wa}.
	\item Type III ("Big Freeze") : In this case, the derivative of the scale factor from the first derivative on wards diverges. These were detected in generalized Chaplygin gas models \cite{bouhmadi2008worse}.
	\item Type IV ("Generalized sudden singularities"): These are finite time singularities with finite density and pressure instead of diverging pressure. In this case, the derivative of the scale factor diverges from a derivative higher than the second \cite{Bamba:2008ut,Nojiri:2004pf}.
	\item Type V ("w-singularities") : In this case, the scale factor, the energy and pressure densities are all finite but the barotropic index $w = \frac{p}{\rho}$ becomes singular \cite{Fernandez-Jambrina:2010ngm}. 
	\item Type $\infty$(" Directional singularities "): Curvature scalars vanish at the singularity but there are causal geodesics along which the curvature components diverge \cite{Fernandez-Jambrina:2007ohv} and in this sense, the singularity is encountered just for some observers. 
	\item Inaccessible singularities: These singularities appear in cosmological models with toral spatial sections, due to infinite winding of trajectories around the
	tori. For instance, compactifying spatial sections of the de Sitter model to cubic tori. However, these singularities cannot be reached by physically well defined observers and hence this prompts the name inaccessible singularities \cite{mcinnes2007inaccessible}.
	\item Type -1 ("Grand Bang/Rip") : In this case, the scale becomes null or diverges for $ w = -1 $ \cite{Fernandez-Jambrina:2014sga}. 
\end{itemize}
\begin{figure}[h]
	\centering
	\includegraphics[width=13cm, height=9cm]{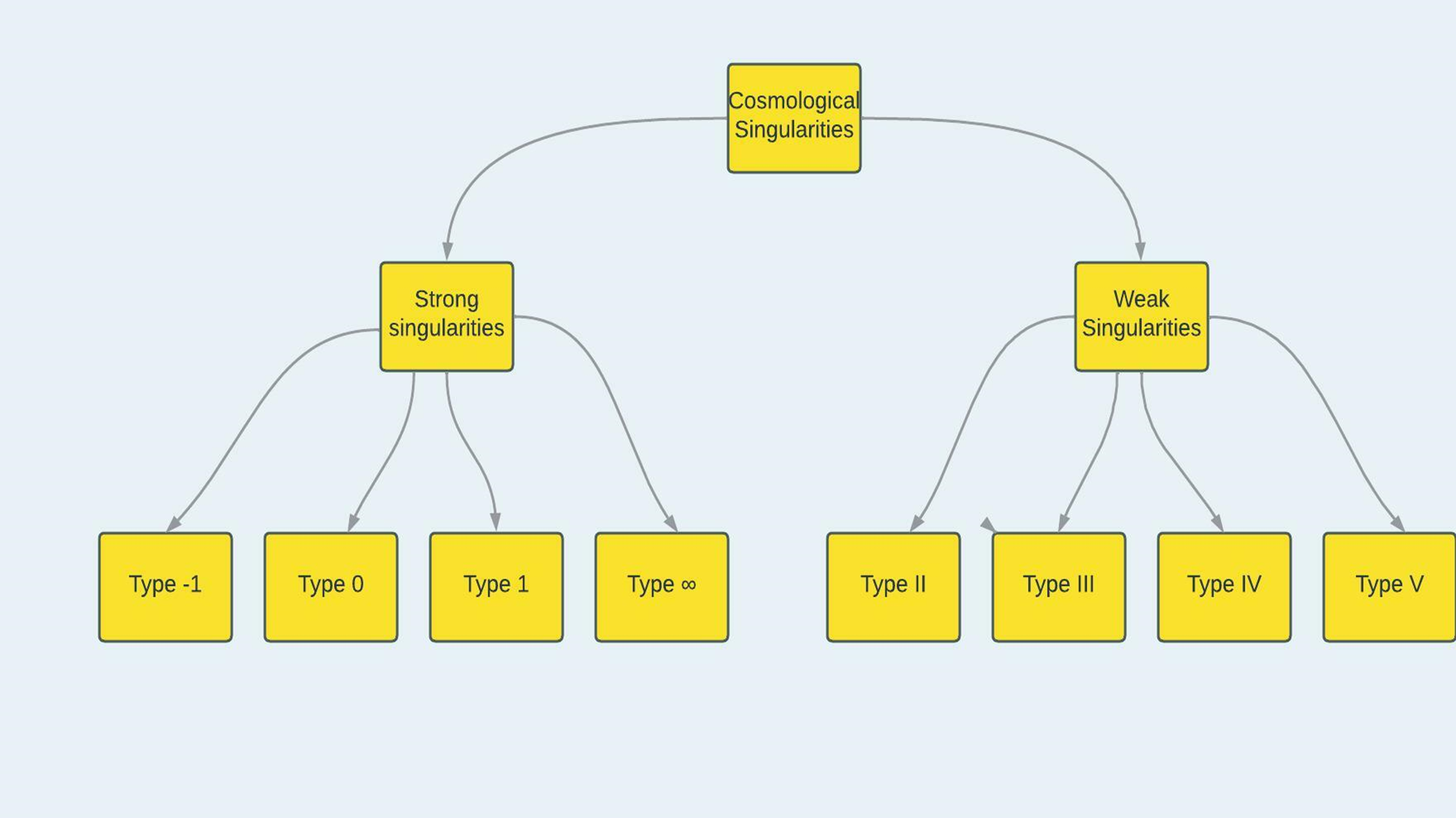}
	\caption{The classification of cosmological singularities summarized}
\end{figure}
All of these singularities discussed above have been studied in a variety of different contexts and in this review, we would like to summarize works primarily of the past two decades on these topics and discuss the current status quo of such singularities. In section II, we would discuss in detail all the singularities we have listed above and how these have been shown to form in different cosmological scenarios. In Section III, we will discuss some lesser known singularities which are more special cases of the previously listed singularities while in section IV, we would discuss various ways which have been shown to remove such singularities (in some cases). In Section V we would discuss a particular dynamical system analysis method (known as the Goriely-Hyde method ) which has been shown to be very useful for cosmological singulartity discussions. Finally in setion VI, we would summarize our brief review and discuss the future outlooks for cosmology with regards to singularities. 
\section{An overview on space-time singularities}
After Einstein proposed the general theory of relativity, which describes gravity in terms of spacetime curvature, the field equations were introduced to relate the geometry of spacetime to the matter content of the universe. Early solutions included the Schwarzschild metric and the Friedmann models. These models described the gravitational field around isolated objects and the overall geometry of the universe, respectively. These models exhibited spacetime singularities where curvatures and energy densities became infinitely high, leading to a breakdown of the physical description. The Schwarzschild singularity at the center of symmetry could be eliminated by a coordinate transformation, but the genuine curvature singularity at \(r=0\) remained. It was initially believed that these singularities were a result of the high symmetry in the models.
\\
\\
However, further research by Hawking, Penrose, Geroch, and others demonstrated that spacetimes could have singularities under more general conditions. Singularities are an inherent feature of the theory of relativity and also apply to other gravitational theories based on spacetime manifolds. These singularities indicate super ultra-dense regions in the universe where physical quantities become infinitely large.In classical theories of gravity, singularities are an unavoidable aspect of describing physical reality. The behavior of these regions is beyond the scope of classical theory, and a quantum theory of gravity is needed to understand them.The field of gravitational physics saw significant developments in the 1960s due to observations of high-energy astrophysical phenomena and advancements in the study of spacetime structure and singularities. These advancements led to progress in black hole physics, relativistic astrophysics, and cosmology.
\\
\\
Singular behavior is observed in space-time models described by general relativity. Examples include the Friedmann-Robertson-Walker (FRW) cosmological models and the Schwarzschild space-time. These models exhibit singularities where energy density and curvatures become infinitely large, leading to a breakdown of the conventional description of space-time.The Schwarzschild space-time displays an essential curvature singularity at \(r=0\), where the Kretschmann scalar \(\alpha=R^{ijkl}R_{ijkl}\) diverges along any non-spacelike trajectory approaching the singularity. Similarly, for FRW models with \(\rho+3p>0\) at all times (where \(\rho\) is total energy density and \(p\) is pressure), a singularity arises at \(t=0\), representing the origin of the universe. Along past-directed trajectories approaching this singularity, both \(\rho\) and the curvature scalar \(R=R_{ij}R^{ij}\) become infinite. In both cases, past-directed non-spacelike geodesics are incomplete, and these essential singularities cannot be eliminated through coordinate transformations.
\\
\\
These singularities represent profound anomalies in space-time, where the usual laws of physics fail. Geodesic incompleteness implies that a timelike observer will cease to exist in the space-time after a finite amount of proper time.While singular behavior can occur without extreme curvature, such cases are considered artificial. An example is the Minkowski space-time with a removed point, where timelike geodesics encounter the hole and become future incomplete. However, it is desirable to exclude such situations by requiring the space-time to be "inextendible," meaning it cannot be isometrically embedded into a larger space-time as a proper subset.
\\
\\
Nevertheless, non-trivial examples of singular behavior exist, such as conical singularities. These singularities do not involve diverging curvature components but are characterized by a Weyl-type solution. An example is the metric given by \(ds^2= -dt^2+dr^2+r^2(d\theta^2+\sin^2\theta\, d\phi^2)\) with the identification \(\phi=0\) and \(\phi=a\) (with \(a\ne 2\pi\)), creating a conical singularity at \(r=0\).The fundamental question is whether such singularities persist in general models and under what conditions they arise. Precisely defining a singularity in a general space-time reveals that singularities likely exist in a broad range of space-times, subject to reasonable conditions. These singularities can emerge as the endpoint of gravitational collapse or in cosmological scenarios, such as the origin of the universe.
\\
\\
The initial observation to make here is that, by its very definition, the metric tensor must possess a well-established meaning at every typical point within the spacetime. However, this principle ceases to hold at a spacetime singularity, like those previously discussed. Such a singularity cannot be considered a standard point within the spacetime; instead, it is a boundary point connected to the manifold. Consequently, difficulties arise when attempting to characterize a singularity based on the requirement that curvatures become infinite in proximity to it. The issue stems from the fact that, since the singularity lies outside the spacetime domain, it is not feasible to define its vicinity in the usual sense, which is essential for discussing the behavior of curvature quantities in that specific region.
\\
\\
An alternative approach might involve defining a singularity in relation to the divergence of elements within the Riemann curvature tensor along trajectories that do not follow spacelike directions. However, a challenge arises here as well: the behavior of these elements can change depending on the reference frames employed, rendering this approach less useful. One might consider utilizing curvature scalars or scalar polynomials involving the metric and Riemann tensor, demanding that they reach exceedingly large values. Instances of such divergence are encountered in models such as Schwarzschild and Friedmann. However, it remains possible that such a divergence only occurs at infinity for a given nonspacelike path. In a broader sense, it seems reasonable to expect some form of curvature divergence to occur along nonspacelike trajectories that intersect a singularity. Nevertheless, attempting to universally characterize singularities through curvature divergence encounters various complications.
\\
\\
Taking into account these scenarios and analogous ones, the presence of nonspacelike geodesic incompleteness is widely accepted as a criterion indicating the existence of a singularity within a spacetime. Although this criterion may not encompass all potential forms of singular behavior, it is evident that the occurrence of incomplete nonspacelike geodesics within a spacetime manifold signifies definite singular behavior. This manifests when a timelike observer or a photon abruptly vanishes from the spacetime after a finite interval of proper time or a finite value of the affine parameter. The singularity theorems, which emerge from an analysis of gravitational focusing and the global attributes of a spacetime, establish this incomplete nature for a broad array of spacetimes under a set of relatively general conditions.
\\
\\
From a physical standpoint, a singularity in any physics theory typically indicates that the theory becomes invalid either in the vicinity of the singularity or directly at the singularity. This implies a need for a broader and more comprehensive theory, necessitating a revision of the existing framework. Similar reasoning applies to spacetime singularities, suggesting that a description involving quantum gravity is warranted within these regions of the universe, rather than relying solely on a classical framework.
\\
\\
The existence of an incomplete nonspacelike geodesic or an inextendible nonspacelike curve with a finite length, as measured by a generalized affine parameter, implies the presence of a spacetime singularity. The concept of "generalized affine length" for such a curve is defined as :$$ L(\lambda) = \int_0^a \left[ \sum_{i=0}^3 (X^i)^2 \right]^{1/2} ds$$
which remains finite. The components $X^i$ represent the tangent to the curve in a tetrad frame propagated in parallel along the curve. Each incomplete curve defines a boundary point of the spacetime, which is singular.To be considered a genuine physical singularity, it is expected that such a singularity is associated with unbounded growth of spacetime curvatures. If all curvature components and scalar polynomials involving the metric and Riemann curvature tensor remain finite and well-behaved as the singularity is approached along an incomplete nonspacelike curve, the singularity might be removable by extending the spacetime with relaxed differentiability requirements \cite{clarke1985conditions}.
\\
\\
Different formalizations are possible for this requirement. A "parallely propagated curvature singularity" is one where the components of the Riemann curvature tensor are unbounded in a parallely propagated frame, forming the endpoint of at least one nonspacelike curve. Conversely, a "scalar polynomial singularity" occurs when a scalar polynomial involving the metric and Riemann tensor takes on infinitely large values along a nonspacelike curve ending at the singularity. This includes cases like the Schwarzschild singularity, where the Kretschmann scalar $(R^{ijkl}R_{ijkl}$ becomes infinite as $r$ approaches 0.Curvature singularities, as further elucidated, also arise in various spacetime scenarios involving gravitational collapse. The strength of singularities and their potential to cause tidal forces on extended bodies can be assessed, and various criteria are available to determine this aspect \cite{ellis1977singular}. These criteria all involve representing a finite object at each point along a causal geodesic as a volume defined by three independent Jacobi fields in the hyperspace, with the velocity of the curve as the normal vector. Tipler's criterion \cite{tipler1977singularities} deems a singularity as strong if this volume tends to zero as the singularity is approached along the geodesic. On the other hand, Krolak's criterion \cite{krolak1986towards} stipulates that the derivative of this volume with respect to the normal parameter must be negative. Consequently, some singularities can be strong according to Krolak's criterion while being weak according to Tipler's, such as type III or Big Freeze singularities. Another criterion is outlined in \cite{rudnicki2006generalized}.

Working with Jacobi fields can be demanding as it involves solving the Jacobi equation along geodesics. Nevertheless, conditions for lightlike and timelike geodesics, satisfying both criteria, have been established \cite{clarke1985conditions}. These conditions are expressed in terms of integrals of the Ricci and Riemann curvatures of the spacetime metric along these curves: \begin{itemize}
	\item Lightlike geodesics: According to Tipler's criterion, a singularity is strong along a lightlike geodesic if and only if the integral
	\[
	\int_{0}^{\tau}d\tau'\int_{0}^{\tau'}d\tau''R_{ij}u^{i}u^j
	\]
	diverges as the normal parameter $\tau$ approaches the singularity.
	Krolak's criterion states that the singularity is strong if and only if the integral
	\[
	\int_{0}^{\tau}d\tau'R_{ij}u^{i}u^j
	\]
	diverges as $\tau$ approaches the singularity.
	\item Timelike geodesics: For timelike geodesics, \cite{clarke1985conditions} presents various necessary and sufficient conditions, but not a single characterization.
	
	Adhering to Tipler's criterion, a singularity is strong along a timelike geodesic if the integral
	\[
	\int_{0}^{\tau}d\tau'\int_{0}^{\tau'}d\tau''R_{ij}u^{i}u^j
	\]
	diverges on approaching the singularity.
	
	Conforming to Krolak's criterion, the singularity is strong if the integral
	\[
	\int_{0}^{\tau}d\tau'R_{ij}u^{i}u^j
	\]
	diverges on approaching the singularity.
	Additional necessary conditions exist, although they are not utilized for our purposes.
\end{itemize}
In passing, it is also of interest to talk of the cosmic censorship conjecture \cite{penrose1969gravitational}, which is the idea that all singularities arising from gravitational collapse will always be hidden by an event horizon. There are actually two versions of this conjecture ; The weak version is that  dynamical singularities in general relativity are generically not visible to observers at infinity, while the strong version is that  dynamical singularities in general relativity are generically not visible to any observer. Singularities in violation of the weak version are dubbed  globally naked, while those in violation of the strong version are dubbed locally naked. The conjectures have not yet been proven and have been a topic of recurring debates and has sprung up a lot of work the topic of naked singularities. In principle, one could think of cosmological singularities as naked singularities as well given that there is no need of an event horizon in such cases and there is no development of such horizon in cases in which such singularities develop too. Several examples of spacetimes containing naked singularities have been found in recent times \cite{joshi2014spacetime,christodoulou1984violation,christodoulou1994examples,joshi1992strong,kuroda1984naked,rodnianski2019naked,Joshi:1993zg,Joshi:2011rlc,Shaikh:2018lcc,Vagnozzi:2022moj,Singh:1994tb,Joshi:2001xi,Joshi:2011zm}. When such singularities develop in gravitational collapse, they give rise again to extremely
intriguing physical possibilities and problems. The opportunity offered in that case is, we may have the possibility to observe the possible ultra-high energy physical processes occurring in such a region of universe,
including the quantum gravity effects. In fact, Loop quantum gravity in particular has a very amicable view of naked singularities and has been shown to be in favour of their existence \cite{Goswami:2005fu,Goswami:2006ph}. Such observations of ultra-high energy events in the universe could
provide observational tests and guide our efforts for a possible quantum theory of gravity and so naked singularities can also be a good avenue for testing out the predictions of such theories. Very recently another very interesting work has been done with regards to naked singularities was performed in \cite{Saurabh:2023otl}. The authors there a performed general relativistic ray-tracing and radiative transfer simulations to generate synchrotron
emission images utilising thermal distribution function for emissivity and absorptivity. They investigate effects in the images of JMN-1 naked singularity and Schwarzschild black hole by varying inclination angle, disk width and frequency. Their results give further motivation of naked singularities being a realistic scenario. 
\\
\\
\section{Types of singularities}
\subsection{Strong singularities }
As mentioned before, strong singularities are those singularities which can distort finite objects in space-time and now we would like to discuss of the prominent singularities in this regard. 
\subsubsection{Big Bang singularity(Type 0) }
Classical models of the universe generically feature an initial or ‘big bang’ singularity. 
This is when we consider progressively earlier and earlier stages of the universe, observable quantities
stop behaving in a physically reasonable way. A more precise mathematical characterisation of
the cosmic big bang singularity can be made in terms of both a global notion of incompleteness
of inextendible causal (i.e., non-spacelike) past-directed curves and a local notion of the existence
of a curvature pathology. Models of inflation also feature massive moving particles seeing a singularity in a finite proper time. The idea of the big bang hence, as is popularly known is, the singularity at the very beginning of the universe. 
\subsubsection{Big rip singularity (Type 1) }
In the case of a big rip singularity, the scale factor of the universe becomes infinite at a finite time in the future, the energy density and the pressure density of the universe also becomes infinite. In a big rip singularity, the dark energy density becomes so large that it causes the expansion of the universe to accelerate at an ever-increasing rate. As a result, the scale factor of the universe increases without bound, and the universe becomes infinitely large at the time of the big rip. The energy density and pressure density of the universe also become infinite at the time of the big rip.The thing to note here is that interestingly the big rip was proposed as a possible phantom scenario for the universe \cite{Caldwell:2003vq}, which means that the equation of state $$w = \frac{p}{\rho} < -1 $$
The phantom conclusion is interesting from the point of view that this presents some peculiar properties, like the energy density of phantom energy increasing with time or the fact that a phantom scenario violates the dominant energy condition \cite{carroll2003can}. Despite the fact that sound waves in quintessence travel at the speed of light, it should not be automatically assumed that disturbances in phantom energy must propagate faster than the speed of light. Indeed, there exist several scalar-field models for phantom energy where the sound speed is actually subluminal \cite{Parker:1999td,Armendariz-Picon:1999hyi,Chiba:1999ka,Faraoni:2001tq}. Phantom constructions have also been discussed in the context of quantum gravitational theories, for example in various string theoretic realizations of dark energy.  \cite{Frampton:2002tu,Sahni:2002dx,McInnes:2001zw}. So it seems in principle interesting to look for a late time universe scenario with phantom dominance and that is where big rip comes in. 
\\
\\
It is also worth discussing the subtleties of the big rip and how it would unfold.In a universe resembling one with a cosmological constant, the scale factor's expansion is faster than the Hubble distance, leading galaxies to gradually vanish beyond our observable horizon. If we introduce the concept of phantom energy, the expansion rate (Hubble constant) increases over time, causing the Hubble distance to shrink. Consequently, galaxies disappear at an accelerated pace as the cosmic horizon approaches. What's even more intriguing is the potential of enhanced dark energy density to eventually tear apart objects held together by gravity. In the framework of general relativity, the gravitational potential's source stems from the volume integral of the sum of energy density ($\rho$) and three times the pressure ($p$), denoted as $\rho+3p$.

For instance, a planet in orbit around a star with mass $M$ and radius $R$ becomes unbound approximately when the condition $-(4\pi/3)(\rho+3p)R^{3} \approx M$ is satisfied. In cases where the equation $-(\rho+3p)$ decreases over time due to a parameter $w$ greater than or equal to -1, if $-(4\pi/3)(\rho+3p) R^{3}$ is smaller than $M$ at present, it will continue to remain smaller indefinitely. This implies that any currently gravitationally bound system, such as the solar system, the Milky Way, the Local Group, and galaxy clusters, will remain bound in the future.
\\

However, when dealing with phantom energy, the quantity $-(\rho+3p)$ increases with time. Consequently, at a certain point in time, every gravitationally bound system will eventually disintegrate. Analyzing the time evolution of the scale factor and the dependence of phantom-energy density on time, we deduce that a gravitationally bound system with mass $M$ and radius $R$ will undergo disintegration around a time $t \approx P \sqrt{2|1+3w|}/[6\pi |1+w|]$. Here, $P$ represents the period of a circular orbit at radius $R$ within the system. This process occurs prior to the Big Rip, with the earliest estimate of the Big Rip time being 35 billion years. Big rip even rips apart molecules, atoms and even nuclei are dissociated ( which makes it fitting that the name of the singularity is the big rip). However, it is not all gloomy as various works have explored way to avoid the big rip too (for example \cite{Elizalde:2004mq}) and we will be discussing those later on here. A lot more works have been done on various aspects of big rip singularities over the years, see \cite{Stefancic:2004kb,Sami:2005zc,Setare:2008pc,Wang:2009rw,Bouhmadi-Lopez:2004mpi,Sami:2006wj,Dabrowski:2006dd,Capozziello:2009hc,Gonzalez-Diaz:2004iks,Astashenok:2012tv,Barrow:2009df,Bamba:2012ka,Gonzalez-Diaz:2004xgp}. The comparison of Big Rip from Dark energy and modified gravity was done for the first time in \cite{nojiri2010future}. Less drastic variants of the big rip have also been found in recent years \cite{frampton2012models,frampton2012pseudo,Wei:2012ct} which have been discussed in detail in Appendix B.
\subsubsection{Grand bang and Grand rip singularities (Type $-1$)}
The grand bang, although apparently different in name, is quite the same as the big bang singularity with null scale factors and a diverging pressure and energy density with the one difference being that the singularity occurs with the equation of state parameter being equal to -1 \cite{Fernandez-Jambrina:2014sga}. This type of singularity was found initially by using a series ansatz for the scale factor.
An understanding of the grand bang and grand rip singularities is quite intricately linked with each other and so we shall discuss that now. 
\\
\\
To discuss these grand singularities, we note that the equation for the parameter \(w\) is given by:
\[ w = \frac{p}{\rho} = -\frac{1}{3} - \frac{2}{3} \frac{a\ddot{a}}{\dot{a}^2}. \]

This expression holds true specifically for flat models. When considering curvature, additional terms need to be included.

The equation of state (EOS) parameter \(w\) has a close connection with the deceleration parameter \(q\):
\[ q = -\frac{a\ddot{a}}{\dot{a}^2} = \frac{1 + 3w}{2}, \]
assuming flat models. Otherwise, the relationship between these parameters becomes more intricate, involving the Hubble parameter \(H = \dot{a}/a\). This enables a direct translation of results from the EOS parameter to the deceleration parameter.

Alternatively, one can view this equation as the differential equation governing the evolution of the scale factor for a given time-dependent barotropic index \(w(t)\). It is advantageous to introduce the variable \(x = \ln a\):
\[ \frac{\ddot{x}}{\dot{x}^2} = -\frac{3}{2}(w + 1) = -(q + 1), \]
allowing us to define
\[ h(t) := \frac{3}{2}(w(t) + 1) = q(t) + 1 \]
as a correction around the case of a pure cosmological constant:
\[ w(t) = -1 + \frac{2}{3}h(t), \quad q(t) = -1 + h(t). \]

This change of variables assists in reducing the order of the differential equation:
\[ h = -\frac{\ddot{x}}{\dot{x}^2} = \left(\frac{1}{\dot{x}}\right)^\cdot \Rightarrow \dot{x} = \left(\int h\,dt + K_{1}\right)^{-1}, \]
In terms of this, one finds the scale factor to be \begin{equation} \label{scale}
a(t)=\exp\left(\int\frac{dt}{\int h(t)\,dt}\right),
\end{equation}
If one then assumes a power series form of $h(t)$ ( which has become quite a well motivated ansatz for the scale factor in various cosmological studies, see ) \begin{equation}
h(t)=h_{0}t^{\eta_{0}}+ h_{1}t^{\eta_{1}}+ \cdots, \qquad 
\eta_{0}<\eta_1<\cdots
\end{equation}
It can then be found out that the energy and pressure densities can be written as 
\[\rho(t)=\left\{\begin{array}{ll}\displaystyle 3\left(\frac{\eta_{0}+1}{h_{0}}\right)^2t^{-2(\eta_{0}+1)}+\cdots
&\textrm{if\ }-1\neq \eta_{0}\neq 0\\\\\displaystyle 
\frac{3}{ h_{0}^2}\frac{1}{\ln^{2}|t|}+\cdots&\textrm{if\ }\eta_{0}=-1
\\\\\displaystyle
\frac{3t^{-2}}{h_{0}^2}
+\cdots 
&\textrm{if\ }\eta_{0}= 0,\end{array}\right.\]
and the pressure,
\[p(t)=\left\{\begin{array}{ll}\displaystyle
\frac{2(\eta_{0}+1)^2}{h_{0}}t^{-\eta_{0}-2}+\cdots 
&\textrm{if\ }-1\neq\eta_{0}<0\\\\\displaystyle 
\frac{2}{ h_{0}}\frac{1}{t\ln^{2}|t|}+\cdots &\textrm{if\ 
}\eta_{0}=-1\\ \\\displaystyle
\frac{2h_{0}-3}{h_{0}^2}t^{-2}
+\cdots  &
\textrm{if\ }\eta_{0}=0\\
\\ -\displaystyle
3\left(\frac{\eta_{0}+1}{h_{0}}\right)^2t^{-2(\eta_{0}+1)}+\cdots &
\textrm{if\ }\eta_{0}>0,
\end{array}\right.\] 
This presents us with intriguing possibilities, where our specific focus will be on the case where $\eta_{0} > 0$. In this scenario, we observe that at $t=0$, $\rho$ and $p$ exhibit divergences following $t^{-2(\eta_{0}+1)}$, and the parameter $w$ converges to the value of $-1$. The consideration of such a singularity has not been explored within previous frameworks. The reason behind this omission is rooted in its incompatibility with the classifications established in \cite{Fernandez-Jambrina:2006tkb} and \cite{Cattoen:2005dx}. This is due to the behavior of the scale factor (which is an exponential of rational functions); it doesn't lend itself to convergent power expansions, whether generalized or not, with a finite number of terms featuring negative powers. However, the function $x(t)$ does exhibit such behavior. The nature of the singularity is governed by the sign of the coefficient $h_{0}$. This is evident in the approximation of $a(t)$ as
\[a(t) \approx e^{-\mathrm{sgn}\,(h_{0})\alpha/t^{\eta_{0}}}, \quad \alpha = \frac{\eta_{0}+1}{\eta_{0}|h_{0}|} > 0, \quad t > 0,\]
Based on this, we make the following observations: \begin{itemize}
	\item For $h_{0} > 0$: In this scenario, the exponential term in equation \eqref{scale} decreases as $t$ increases, and the scale factor $a$ approaches zero as $t$ approaches 0. This resembles an exponential-type Big Bang singularity or, if we swap $t$ for $-t$, a Big Crunch. Given that $h_{0}$ is positive, the barotropic index $w$ consistently remains below the phantom divide near $t=0$. Specifically, the value $w=-1$ is approached from values below it. These types of singularities are known as grand bang singularities.
	\item For $h_{0} < 0$: Conversely, in this case, the exponential term increases as $t$ increases, causing the scale factor $a$ to diverge to infinity as $t$ approaches 0. This resembles an exponential-type Big Rip singularity at $t=0$, which, when considering the future, can be located by substituting $t$ with $-t$. In this instance, the barotropic index $w$ consistently remains above the phantom divide, and the value $w=-1$ is approached from values above it. This scenario is termed the grand rip singularity.
\end{itemize}
\subsubsection{Directional singularities (Type $\infty$)} 
We follow the discussion of \cite{Fernandez-Jambrina:2007ohv} in order to understand how directional singularities were initially found in cosmological models. If we start with the flat FLRW metric in the form indicated by Eq. (1):
\begin{equation}\label{metric}
ds^2=-dt^2+a^2(t)\left(dr^2+ r^2\left(d\theta^2+\sin^2\theta d\phi^2\right)\right)
\end{equation}
it is evident that the equations governing the trajectories of geodesics, followed by observers not subject to acceleration ($\delta=1$) and light-like particles ($\delta=0$) possessing a specific linear momentum $P$, can be simplified to:
\begin{equation} \label{geods1}
\frac{dt}{d\tau}= \sqrt{\delta
	+\frac{P^2}{a^2(t)}}
\end{equation}
\begin{equation}\label{geods2}
\frac{dr}{d\tau}=\pm\frac{P}{a^2(t)}
\end{equation}
assuming constant $\theta$ and $\phi$ due to the symmetry inherent in these models. Here, $\tau$ represents the intrinsic or proper time as measured by the observer.In the context of null geodesics, we find that: \begin{equation}
\Delta\tau=\frac{1}{P}\int_{-\infty}^ta(t) dt
\end{equation}
Consequently, to ensure that the initial \emph{event} $t=-\infty$ corresponds to a finite proper time interval $\Delta\tau$ from an event at $t$, the requirement is:
\begin{equation} \label{condit}
\int_{-\infty}^ta(t)\,dt<\infty.
\end{equation}
Therefore, the emergence of singular behavior exclusively at $t=-\infty$ is possible if the scale factor can be expressed as an integrable function of coordinate time. This condition necessitates that $a(t)$ tends towards zero as $t$ approaches $-\infty$, although this alone is not sufficient.
Similarly, for timelike geodesics with non-zero $P$: \begin{equation}
\Delta \tau=\int_{-\infty}^t\frac{dt}{\sqrt{1+\frac{P^2}{a^2(t)}}}<\frac{1}{P}\int_{-\infty}^t a(t) dt
\end{equation} indicating that the proper time interval to $t=-\infty$ is finite provided the time interval for light-like geodesics is also finite. Consequently, $t=-\infty$ is reachable for these observers.
As a result, condition \eqref{condit} implies that both light-like and timelike geodesics with non-zero $P$ experience $t=-\infty$ within a finite proper time interval in their past.
Conversely, comoving observers tracing timelike geodesics with $P=0$ exhibit $d\tau=dt$, which leads to $t=-\infty$ corresponding to an infinite proper time interval in their past, and thus, they cannot encounter the singularity.
This dichotomy is responsible for the directional nature of Type $\infty$ singularities, as they are accessible to causal geodesics except those with $P=0$. Ultimately, it can be concluded that Type $\infty$ singularities can manifest in three scenarios: \begin{itemize}
	\item For a finite $\int_{-\infty} h\,dt$ with $h(t)>0$: $a_{-\infty}=0$,
	$\rho_{-\infty}=\infty$, $p_{-\infty}=-\infty$, $w_{-\infty}=-1$. These differ from the "little rip" model in the sign of $h(t)$, and are termed "little bang" if they denote an initial singularity, or "little crunch" if they represent a final singularity \cite{Fernandez-Jambrina:2016clh}. Instances of this case encompass models with a scale factor $a(t)\propto e^{-\alpha(-t)^p}$ where $p>1$ and $\alpha>0$.
	\item When $h_{-\infty}=0$ and $|h(t)|\gtrsim |t|^{-1}$ with $h(t)<0$: $a_{-\infty}=0$, $\rho_{-\infty}=0$, $p_{-\infty}=0$, $w_{-\infty}=-1$. Changing the sign of $h(t)$ gives rise to a variant of the "little rip" scenario, featuring an asymptotically vanishing energy density and pressure. Models with a scale factor $a(t)\propto e^{-\alpha(-t)^p}$ where $p\in(0,1)$ and $\alpha>0$ exemplify this case.
	\item  For a finite $h_{-\infty}\in(-1,0)$: $a_{-\infty}=0$, $\rho_{-\infty}=0$,
	$p_{-\infty}=0$, and a finite $w_{-\infty}\neq-1$. This case applies to models like $a(t)\propto t^{-p}$ with $p>1$, as explored in \cite{Fernandez-Jambrina:2007ohv}. 
\end{itemize}
While they have recently been discussed in the context of inflationary models \cite{Fernandez-Jambrina:2016clh},not much work has been done in the regard of Type $\infty$ singularities since their discovery with regards to their avoidance or their emergence in more exotic cosmological models.
\subsection{Weak singularities}
\subsubsection{Sudden singularities (Type II) }
In the case of such type II singularities, one has the pressure density diverging or equivalently, the derivatives of the scale factor diverging from the second derivative onwards. Let's start by examining informally whether there is a potential for the emergence of singularities in which a physical scalar quantity becomes unbounded at a finite future comoving proper time \(t_{s}\). This might occur when the scale factor \(a(t)\) approaches a non-zero or infinite value \(a(t_{s})\) and the Hubble parameter \(H(t)\) approaches a finite value \(H_{s}\) (where \(H_{s}\) is positive and not infinite). If such scenarios are feasible, the following conditions need to be satisfied:
\begin{equation}
\rho \rightarrow 3H_{s}^{2}+\frac{k}{a_{s}^{2}}=\rho _{s}<\infty
\label{lim1}
\end{equation}

and

\begin{equation}
\frac{\ddot{a}}{a}
\rightarrow \frac{p}{2}\ -\frac{H_{s}^{2}}{2}-\frac{k}{6a_{s}^{2}}
\label{lim2}
\end{equation}

\begin{equation}
\dot{\rho}\rightarrow -3H_{s}(\rho _{s}+p)  \label{lim3}
\end{equation}

Hence, it becomes apparent that the density must inevitably remain finite at \(t_{s}\). However, there is still a possibility for a singularity in pressure to arise, manifested as:

\begin{equation}
p(t)\rightarrow \infty  \label{p}
\end{equation}

as \(t\rightarrow t_{s}\), consistent with the conditions outlined in Equation \eqref{lim2}. In such instances, the pressure singularity is concomitant with an infinite acceleration. 
\\
\\
To illustrate an example for this, we take the most primitive example of such singularities which was put forward by Barrow in \cite{Barrow:2004xh}. In this regard, assume that it is physically reasonable to expect that the scale factor can be written in the form of this ansatz \footnote{Appendix A provides a detailed overview behind the motivations that allow us to make such a consideration for the scale factor} \begin{equation}
a(t) = A + Bt^{q} + C(t_{s}-t)^{n},  \label{ex}
\end{equation}

where \(A > 0\), \(B > 0\), \(q > 0\), \(C\), and \(n > 0\) are constants that we will determine. We set the origin of time such that \(a(0) = 0\), leading to \(A = -Ct_{s}^{n} > 0\). Consequently, we find the expression for the Hubble parameter \(H_{s}\):

\begin{equation}
H_{s} = \frac{qBt_{s}^{q-1}}{A + Bt_{s}}.  \label{H}
\end{equation}

For simplicity, we use the freedom to scale the Friedmann metric by dividing by \(A\) and set \(A \equiv 1\) and \(C \equiv -t_{s}^{n}\). This yields the simplified form of \(a(t)\):

\begin{equation}
a(t) = \left(\frac{t}{t_{s}}\right)^{q} (a_{s} - 1) + 1 - \left(1 - \frac{t}{t_{s}}\right)^{n},
\label{sol2}
\end{equation}

where \(a_{s} \equiv a(t_{s})\). As \(t\) approaches \(t_{s}\) from below, the behavior of the second derivative of \(a\) can be described:

\begin{equation}
\ddot{a} \rightarrow q(q-1)Bt^{q-2} - \frac{n(n-1)}{t_{s}^{2}(1-\frac{t}{t_{s}})^{2-n}} \rightarrow -\infty,
\label{Lim}
\end{equation}

whenever \(1 < n < 2\) and \(0 < q \leq 1\). This solution is valid for \(0 < t < t_{s}\). Consequently, as \(t\) approaches \(t_{s}\), \(a\) approaches \(a_{s}\), \(H_{s}\) and \(\rho_{s} > 0\) (as long as \(3q^{2}(a_{s}-1)^{2}t_{s}^{-2} > -k\)) remain finite, while \(p_{s} \rightarrow \infty\).

When \(2 < n < 3\), \(\ddot{a}\) remains finite but \(\dddot{a} \rightarrow \infty\) as \(t\) approaches \(t_{s}\). Here, \(p_{s}\) remains finite, but \(\dot{p}_{s} \rightarrow \infty\).In contrast, there exists an initial strong-curvature singularity, where both \(\rho\) and \(p\) tend to infinity as \(t\) approaches 0. Importantly, in this scenario, both \(\rho\) and \(\rho + 3p\) remain positive. Such behavior can even arise in a closed universe (\(k = +1\)), where the pressure singularity prevents expansion from reaching a maximum. This is the most primitive example of a pressure singularity but ever since the work in \cite{Barrow:2004xh}, such singularities have been discussed in loads of various different settings, both from modified gravity perspectives and other phenomenological considerations. Work has also been done on ways to escape such singularities and we will be discussing them later in our discussions. 
\subsubsection{Big Freeze singularity (Type III) }
The Big Freeze singularity is similar to the Big rip but is still quite different from it. This singularity was firstly shown in a phantom generalized Chaplygin gas cosmology (PGCG) in \cite{bouhmadi2008worse} and we shall quickly see how it unfolds in such a scenario.
\\
The equation of state governing PGCG closely resembles that of the conventional generalized Chaplygin gas. It can be succinctly expressed as:

\[ p = -\frac{A}{\rho^{\alpha}}, \]

where the symbol \( A \) represents a positive constant and \( \alpha \) signifies a parameter. In the scenario where \( \alpha = 1 \), the equation assumes the form of a simple Chaplygin gas equation of state. This relationship is crucially connected to the continuity equation, given by:

\[ \dot{\rho} + 3 H (p + \rho) = 0, \]

from which emerges the expression for energy density \( \rho \):

\[ \rho = \left(A + \frac{B}{a^{3(1+\alpha)}}\right)^{\frac{1}{1+\alpha}}, \]

where \( B \) stands as a constant parameter. In a noteworthy observation made in Ref.~\cite{Bouhmadi-Lopez:2004mpi}, it was discerned that a negative \( B \) renders the perfect fluid, with the equation of state \( p = -\frac{A}{\rho^{\alpha}} \), unable to uphold the null energy condition, that is, \( p + \rho < 0 \). Intriguingly, under these conditions, the energy density rather escalates as the Universe expands, contrary to redshift behavior, thus earning the term "phantom generalized Chaplygin gas" (PGCG).

Further insights from the works of \cite{Bouhmadi-Lopez:2004mpi,Sen:2005sk} unveil that for PGCG with \( \alpha > -1 \), a FLRW Universe hosting this fluid can evade the impending big rip singularity. As the scale factors attain magnitudes far beyond, the Universe eventually approximates an asymptotically de Sitter state. In stark contrast, during the Big Freeze scenario, the PGCG energy density responds by amplifying as the scale factor matures. Especially, as the scale factor approaches minuscule values (\( a \rightarrow 0 \)), \( \rho \) tends towards \( A^{\frac{1}{1+\alpha}} \), while it experiences a surge at a finite scale factor \( a_{\rm{max}} \):

\[ a_{\rm{max}} = \left|\frac{B}{A}\right|^{\frac{1}{3(1+\alpha)}}. \]

As a consequence, a FLRW Universe saturated with PGCG is destined to confront a finite-radius future singularity. Notably, the vicinity of this singularity lends itself to a cosmological evolution described by the relation:

\[ a \simeq a_{\rm{max}} \left\{1 - \left[\frac{1+2\alpha}{2(1+\alpha)}\right]^{\frac{2(1+\alpha)}{1+2\alpha}} A^{\frac{1}{1+2\alpha}} |3(1+\alpha)|^{\frac{1}{1+2\alpha}} (t_{\rm{max}} - t)^{\frac{2(1+\alpha)}{1+2\alpha}}\right\}. \]

Remarkably, this singularity not only emerges at a finite scale factor but also at a distinct future cosmic time. Conversely, the history of a FLRW Universe permeated with this fluid traces back to an asymptotically de Sitter state in the past. Expressing this temporal journey succinctly:

\[ a \simeq a_0 \exp\left(A^{\frac{1}{2(1+\alpha)}}t\right), \]

where \( a_0 \) signifies a minute scale factor. Additionally, the Universe embarks on its odyssey from a bygone infinity of cosmic time, as \( a \rightarrow 0 \) and \( p+\rho \rightarrow 0^- \). Remarkably, the homogeneous and isotropic nature of the Universe propels it into a phase of super-accelerated expansion, denoted by:

\[ \dot H = -\frac{3}{2} (p+\rho) > 0, \]

up until it culminates at the singularity \( a = a_{\rm{max}} \). It's imperative to recall that the PGCG eludes satisfaction of the null energy condition \cite{Bouhmadi-Lopez:2004mpi}, as embodied in \( p+\rho < 0 \). A great amount of work has been done on Big Freeze singularities since the intial one by in \cite{bouhmadi2008worse} and it has been shown that one can encounter such singularities in a lot of exotic cosmological settings and also there have been works probing how one can avoid such singularities too \cite{Bouhmadi-Lopez:2014tna,Bouhmadi-Lopez:2014jfa,Wetterich:2014eaa,Houndjo:2012ij,Singh:2010qa,Belkacemi:2011zk,Bouhmadi-Lopez:2009ggt,Yurov:2007tw}. 
\subsubsection{Generalized sudden singularities (Type IV) }
These singularities were firstly discussed in \cite{Nojiri:2005sx} and have since then been found in a diverse variety of cosmological settings. So here we will briefly discuss the primary cases in which type IV singularities were shown. In fact, this example will illustrate all the prominent singularities we have discussed so far. We start with an equation of state of the form  \begin{equation}
p=-\rho - f(\rho)
\end{equation}
This sort of equation of state with $ f(\rho)  = A \rho^{\alpha} $ for $\alpha$ being an arbitrary constant was first proposed in \cite{Nojiri:2004pf} and was investigated in detail in \cite{Stefancic:2004kb} and there can be diverse physical motivations behind such an equation of state. This form of EOS can also be equivalent to bulk viscosity \cite{Barrow:1986yf}. This type of an equation of state can also come about due to modified gravity effects \cite{Nojiri:2005sr}. We now consider the following ansatz for the scale factor 
\begin{equation}
\label{EOS7}
a(t)=a_0 \left(\frac{t}{t_s - t}\right)^n \,.
\end{equation}
where $n$ is a positive constant and $0 < t < t_s$. The scale factor diverges within a finite time ($t \to t_s$), resembling the phenomenon of the Big Rip singularity. Consequently, $t_s$ represents the universe's lifetime. When $t \ll t_s$, the evolution of $a(t)$ follows $t^n$, leading to an effective EOS given by $w = -1 + 2/(3n) > -1$. Conversely, when $t \sim t_s$, the effective EOS assumes $w = -1 - 2/(3n) < -1$. The Hubble rate in this case can be expressed as
\begin{equation}
\label{eq:hubble_rate}
H = n \left(\frac{1}{t} + \frac{1}{t_s - t}\right)\,.
\end{equation}
Utilizing Equation \eqref{eq:hubble_rate} one can deduce the relation
\begin{equation}
\label{eq:density}
\rho = \frac{3n^2}{\kappa^2} \left(\frac{1}{t} + \frac{1}{t_s - t}\right)^2\,.
\end{equation}
As a result, both $H$ and $\rho$ exhibit minima at $t = t_s/2$, characterized by the values
\begin{equation}
\label{eq:minima_values}
H_{\text{min}} = \frac{4n}{t_s}\ ,\quad
\rho_{\text{min}} = \frac{48n^2}{\kappa^2 t_s^2}\ .
\end{equation}
Next, we examine a specific form for $f(\rho)$ given by
\begin{equation}
\label{eq:f_rho}
f(\rho) = \frac{AB \rho^{\alpha+\beta}}{A\rho^\alpha + B \rho^\beta}\
\end{equation}
where $A$, $B$, $\alpha$, and $\beta$ are constants. As we shall see, this dark energy scenario harbors a complex structure with respect to singularities.

In scenarios where $\alpha$ surpasses $\beta$, we observe that
\begin{equation}
\label{eq:f_rho_limits}
f(\rho)\to \begin{cases}
A\rho^\alpha & \text{as}\ \rho\to 0 \\
B\rho^\beta & \text{as}\ \rho\to \infty \\
\end{cases} \ .
\end{equation}
For non-unit values of $\alpha$ and $\beta$, we obtain
\begin{equation}
\label{eq:scale_factor_exp}
a = a_0 \exp \left\{-\frac{1}{3}\left[\frac{\rho^{-\alpha + 1}}{(\alpha - 
	1)A}
+ \frac{\rho^{-\beta + 1}}{(\beta - 1)B}\right] \right\}\,.
\end{equation}
The realm of possibilities in this cosmology is extensive. If $1 > \alpha > \beta$ and $A,B > 0$ ($A,B < 0$), the scale factor has a minimum (maximum) at $\rho=0$, extending to infinity (vanishing) as $\rho\to \infty$. When $\alpha>1>\beta$ and $A<0$ while $B>0$ ($A>0$ and $B<0$), the scale factor features a minimum (maximum) at a non-trivial (non-vanishing) $\rho$ value, reaching infinity (zero) as $\rho$ approaches zero or a positive infinity. For $\alpha>1>\beta$ and $A,B > 0$ ($A,B < 0$), the scale factor becomes infinite (vanishes) as $\rho \to \infty$ ($\rho \to 0$), and it vanishes (increases) as $\rho\to 0$ ($\rho\to \infty$). When $\alpha>\beta>1$, the scale factor approaches $a_0$ as $\rho\to \infty$. Additionally, if $A>0$ ($A<0$), the scale factor tends to $0$ ($\infty$) as $\rho\to 0$. With $A,B > 0$ ($A,B < 0$), the scale factor demonstrates monotonic growth (decrease) concerning $\rho$. In the case of $A>0$ and $B<0$ ($A<0$ and $B>0$), the scale factor attains a nontrivial maximum (minimum) at a finite $\rho$ value.
\\
\\
To summarize, the possibilities for singularity formation in this cosmological model are remarkably diverse. It's worth noting that some of the identified singularities may violate one or more energy conditions. These energy conditions encompass:
\begin{align}
&\rho \geq 0 \quad \rho \pm p \geq 0 \qquad \text{"dominant energy condition"} \\
&\rho + p \geq 0 \qquad \text{"null energy condition"} \\
&\rho \geq 0 \quad  \rho + p \geq 0 \qquad \text{"weak energy condition"} \\
&\rho + 3p \geq 0 \quad \rho + p \geq 0 \qquad \text{"strong energy condition"}
\end{align}
With these considerations, we can succinctly summarize the findings for the cosmological model defined by the $f(\rho)$ function as follows:
\begin{itemize}
	\item For $A/B<0$, a type II singularity is inevitable, irrespective of the values of $\beta$.
	\item Regardless of the sign of $A/B$, the nature of singularities varies according to the values of $\beta$.
	\begin{enumerate}
		\item $0<\beta<1/2$: A type IV future singularity is evident. The parameter $w$ approaches infinity ($-\infty$) for $B<0$ ($B>0$).
		\item $\beta>1$: A type III future singularity emerges, accompanied by a breach of the dominant energy condition. The parameter $w$ approaches infinity ($-\infty$) for $B<0$ ($B>0$).
		\item $3/4<\beta<1$: A type I future singularity emerges if $A>0$. The dominant energy condition is violated for $A>0$, and $w$ approaches $-1+0$ ($-1-0$) for $A<0$ ($A>0$).
		\item $1/2 \le \beta \le 3/4$: No finite future singularity is present.
		\item $\beta=0$: A finite future singularity is absent, yet as $\rho \to 0$, $w$ approaches infinity ($-\infty$) for $B<0$ ($B>0$).
		\item $\beta<0$: A type II future singularity emerges. The dominant energy condition is broken, though the strong energy condition remains intact for $B<0$. The parameter $w$ approaches infinity ($-\infty$) for $B<0$ ($B>0$).
	\end{enumerate}
\end{itemize}
So with this example (as was discussed in \cite{Nojiri:2005sx} ) shows us how one can find not only type IV singularities but also the other singularities we have discussed so far. Another interesting thing to note is that there comes out to be qualitative differences when one considers singularities in Jordan and Einstein frames, something which was discussed in detail and discovered in \cite{briscese2007phantom,bahamonde2016correspondence}. It is also worth noting that when one considers viscous fluids, as in \cite{brevik2017viscous}, then it may give rise to different types of singularities. The occurrence of singularities in an oscillating universe has been also been discussed, firstly in \cite{nojiri2006oscillating}. Singularities have also been considered in detail bounce cosmologies \cite{odintsov2015bouncing,odintsov2017big,bamba2011time}. The realisation of all known 4 types of future singularities (Type I-Type IV) has also been found in very exotic modified gravity theories, for example in an f(R) version of Horava-Lifschitz gravity \cite{carloni2010modified}, while also in Teleparallel constructions like the one considered in \cite{bamba2012reconstruction}. 
\\
\\
A crucial point to that we should note here in passing with regards to all the singularities we have discussed so far is that the tidal forces which manifest for these singularities as the (infinite) impulse which reverses (or stops) the increase of separation of geodesics and the geodesics themselves can evolve further ; the universe can continue its evolution through a singularity. Moreover, it's intriguing to consider the potential consequences of these singularities on the constructs of quantum gravity. Although there exists a considerable body of literature exploring the emergence of cosmological singularities in quantum gravitational scenarios like Braneworlds, for instance, a more profound inquiry pertains to the influence of such singularities on fundamental entities like strings.

If we contemplate an elongated structure such as a classical string, modeled using the Polyakov formalism \cite{polchinski1996string}:
\begin{equation}
S = - \frac{T}{2} \int d\tau d\sigma \eta^{\mu\nu}g_{ab} \partial_{\mu}
X^{a} \partial_{\nu} X^{b}
\end{equation}
(with $T$ denoting string tension, and $\tau, \sigma$ representing the string's worldsheet coordinates, $\eta^{\mu\nu}$ corresponding to the worldsheet metric, $\mu,\nu = 0,1$, and $g_{ab}$ standing for the spacetime metric), the scenario involves the string interacting with a non-BB singularity \cite{balcerzak2006strings}. The crux of the matter is that a measurable property of the string, its invariant size $S(\tau) = 2 \pi a(\eta(\tau)) R(\tau)$ (assuming a circular assumption with radius $R$), reveals certain characteristics. Specifically, at a Big-Rip singularity, the string undergoes infinite stretching ($S \to \infty$), resulting in its destruction. In contrast, at a type II singularity, the scale factor remains finite at the $\eta$-time, consequently maintaining a finite invariant string size. Analogously, the same holds true for Type III and Type IV singularities. This implies that strings remain intact when encountering such singularities. \footnote{This also underscores the "weakness" of these singularities in the aspect that they don't display geodesic incompleteness. As a result, particles \cite{fernandez2004geodesic}, and even more extensive entities like extended objects \cite{balcerzak2006strings}, can traverse them without obstruction. Hence, they lack a "dangerous" quality, which explains their potential emergence in the relatively proximate future (for instance, around 10 million years for Type II or the idea that a pressure singularity has happened in the recent past) \cite{denkiewicz2012cosmological,denkiewicz2012observational,Odintsov:2022eqm,Balcerzak:2023ynk}.}
\subsubsection{w singularities (Type V) }
As the name suggests, w singularities occur when the equation of state parameter (w) blows up in some cosmological models. The singularities were firstly introduced in \cite{Dabrowski:2009kg} and then expanded upon in later works \cite{Fernandez-Jambrina:2010ngm,dabrowski2014singularities}. The authors in \cite{Dabrowski:2009kg} arrived at w-singularities by firstly choosing the scale factor ansatz as \begin{equation}
\label{SF1}
a(t)=A+B\left(\frac{t}{t_s}\right)^{\frac{2}{3\gamma}}+C\left(D-\frac{t}{t_s}\right)^n.
\end{equation}
It contains seven arbitrary constants: $A$, $B$, $C$, $D$, $\gamma$, $n$, and $t_s$. The last of the constants $t_s$ is the time when we expect
the singularity. Having the scale factor \eqref{SF1}, they imposed the following conditions
\begin{equation}
\label{cond1}
a(0)=0, \hspace{0.1cm} a(t_s)=const.\equiv a_s, \hspace{0.1cm} \dot{a}(t_s)=0, \hspace{0.1cm} \ddot{a}(t_s)=0~.
\end{equation}
The first of the conditions \eqref{cond1} is chosen in order for the evolution to begin with a standard big-bang singularity at
$t=0$ (note that in order to have a big-rip, one would have to impose $a(0) = \infty$, which is equivalent to taking $\gamma<0$).
One can see that after introducing \eqref{cond1}, the energy density and the pressure
vanish at $t=t_s$. The model does not admit a singularity of the higher derivatives of the Hubble parameter since $\ddot{H}(t_s) \neq 0$ in $\ddot{H}$, and so it is not of the type IV
singularity according to the classification of Ref. \cite{Nojiri:2005sx}. On the other hand, even though both $\ddot{a}(t_s)$ and $\dot{a}(t_s)$ vanish
in the limit $t \to t_s$, the deceleration parameter blows-up to infinity, i.e.,
\begin{equation}
\label{qts}
q(t_s) = - \frac{\ddot{a}(t_s)a_s}{\dot{a}^2(t_s)} \to \infty
\end{equation}
and consequently as one can find the EOS parameter to be related with the deceleration parameter as 
\begin{equation}
w(t) = \frac{c^2}{3}\left[2 q(t) - 1 \right]
\end{equation}
one finds that $w(t_{s}) \to \infty $. Then, we face a very strange singularity. It has vanishing pressure and energy density, a constant scale factor, but the deceleration parameter and, in particular, a time-dependent barotropic index $w(t)$ are singular. Another ansatz for the scale factor which can give w-singularities was proposed by Dabrowski and Marosek in \cite{Dabrowski:2012eb}, which has an exponential form. That ansatz is given by \begin{equation} \label{wansatz}
a(t) = a_{s} \left( \frac{t}{t_{s}}\right)^m \exp \left(1 - \frac{t}{t_{s}}\right)^{n}
\end{equation}
where $a_{s}$ has the units of length and is a constant while m and n are also constants \footnote{While the ansatz on the surface looks quite different from a power series one which we will consider later on, it can be a sub case of a series ansatz within certain limits as well } . The scale factor is zero (a=0) at t=0, thus signifying the big bang singularity. One can write the first and second derivatives of the scale factor as \begin{equation}
\dot{a}(t) = a(t) \left[ \frac{m}{t} - \frac{n}{t_{s}} \left(1 - \frac{t}{t_{s}}\right)^{n-1}  \right] 
\end{equation}
\begin{equation}
\ddot{a}(t)  = \dot{a}(t) \left[ \frac{m}{t} - \frac{n}{t_{s}} \left(1 - \frac{t}{t_{s}}\right)^{n-1}  \right] + a(t) \left[ - \frac{m}{t^2} + \frac{n(n-1)}{t^2} \left(1 - \frac{t}{t_{s}}\right)^{n-2}  \right]
\end{equation}
where the overdots now denote differentiation with respect to time. From this, one can see that for $1<n<2$ $\dot{a} (0) \to \infty $ and $ \dot{a} (t_{s}) = \frac{m a_{s}}{t_{s}} = $ const. ,  while $a(t_{s}) = a_{s} $ , $\ddot{a}(0) \to \infty $ and $ \ddot{a} (t_{s}) \to -\infty $ and we have sudden future singularities. Furthermore, it was shown in \cite{Dabrowski:2012eb} that for the simplified case of the scale factor (20) with $m=0$, one can get w-singularities for $ n > 0 $ and $ n \neq 1 $.
Finally, yet another ansatz to get w-singularities was provided in \cite{Fernandez-Jambrina:2010ngm} and is of a power series form given by \begin{equation}
a(t) = c_{0} + c_{1} (t_{s} - t)^{n1} + c_{2} (t_{s} - t)^{n2}......
\end{equation} 
where $t_{s} $ is the time of the singularity. In order for pressure to be finite, $ n_{1} > 1 $.
There have of course been a significant amount of works which have considered how these singularities can occur in non-standard cosmologies and how they can be avoided too. But in passing, a discussion is in order over the cosmological significance of w-singularities. While Type I-Type IV singularities deal with more direct cosmological parameters like the scale factor, Hubble parameter alongside energy and pressure densities, type V singularities deal with a somewhat indirect parameter in the form of w. This is not to say, however, that these singularities cannot occur in cosmological and in particular, dark energy models. For example \cite{Elizalde:2018ahd}  discussed how w-singularities can occur in interacting dark energy models(while the background cosmology in this case was still general relativistic and the continuity equation had its usual form), while in \cite{Khurshudyan:2018kfk} it was showed how varying Chaplygin gas models can also have w-singularities. The occurrence of w-singularities in various other contexts has also been discussed in \cite{Szydlowski:2017evb,Samanta:2017qnn,Sadri:2018lzz,Astashenok:2012tv,ozulker2022dark}. Hence while type V singularities deal primarily with a more indirect cosmological parameter, it by no means diminishes its cosmological importance and it does appear in a variety of cosmological expansion scenarios.
\section{Singularity removal/avoidance methods}
With huge influx of interest in finding singularities in cosmological models, a natural interest also grew in investigating ways in which such singularities could either be completely removed or at least mildly alleviated/avoided in some cases. This has also resulted in an impressive amount of literature ( for example, look at \cite{capozziello2009classifying} for a detailed account of avoiding singularities in both Jordan and Einstein frames). What we would like to do here is to discuss some of the prominent works in this regard, focusing on the use of quantum effects and modified gravity effects to deal with singularities.
\subsection{Conformal anomaly effects near singularities}
The effect of quantum backreaction of conformal matter around Type I, Type II and Type III singularities were taken into consideration in the works of Nojiri and Odintsov \cite{Nojiri:2005sx,Nojiri:2004ip,Nojiri:2000kz}. In these cases, the curvature of the universe becomes
large around the singularity time $ t= t_{s} $, although the
scale factor a is finite for type II and III singularities.
Since quantum corrections usually contain the powers
of the curvature or higher derivative terms, such correction terms are important near the singularity. At this point, it becomes important to add a bit of context about what conformal anomalies are and how they are usually perceived in high energy physics. It is fair to assume that there are many matter fields during inflation in the early universe because the Standard Model of particle physics has almost 100 fields, and this number may increase by two if the Standard Model is contained in a supersymmetric theory. Although the behaviour of these (massless) matter fields—scalars, the Dirac spinors, and vectors in curved space-time—is conformal invariant, some divergences are observed because of the presence of the one-loop vacuum contributions. In the renormalized action, some counterterms are required to break the matter action's conformal invariance in order to cancel the poles of the divergence component. From the classical point of view, the trace of the energy momentum tensor in a conformally invariant theory is null. But  renormalization procedures can lead to the trace of an anomalous energy momentum tensor, which is the so-called quantum anomaly or the conformal anomaly(we would recommend the reader \cite{deser1976non,duff1994twenty,birrell1984quantum,Bamba:2014jia} for more details on conformal anomaly effects). The conformal anomaly we have described be considered to have the following form \cite{Nojiri:2005sx} \begin{equation}
T_{A} = b \left( F + \frac{2}{3} \Box R \right) + b^{\prime} G + b^{\prime \prime} \Box R
\end{equation} 
where $T_{A}$ is the trace of the stress-energy tensor, F is the square of the 4d Weyl tensor and G is a Gauss-Bonet curvature invariant, which are given by \begin{equation}
F = (1/3) R^2 - 2 R_{i j} R^{i j} + R_{i j k l} R^{i j k l }
\end{equation} 
\begin{equation}
G = R^2 - 4 R_{i j} R^{i j } + R_{i j k l} R^{i j k l}
\end{equation}
b and $ b^{\prime} $ on the other hand are given by \begin{equation}
b = \frac{N + 6 N_{1/2} + 12 N_{1} + 611 N_{2} - 8 N_{HD}}{120 (4 \pi)^2 }
\end{equation} 
\begin{equation}
b^{\prime} = - \frac{N + 11 N_{1/2} + 62 N_{1} + 1411 N_{2} - 28 N_{HD} }{360 (4 \pi)^2 }
\end{equation}
with N scalar, $ N_{1/2} $ spinor , $ N_{1} $ vector fields  , $ N_{2} $ (= 0 or 1 ) gravitons and $ N_{HD} $ being higher derivative conformal scalars. For usual matter, $ b > 0 $ and $ b^{\prime} < 0 $ except for higher derivative conformal scalars while $ b^{\prime \prime} $ can be arbitrary. Quantum effects due to the conformal anomaly act as a fluid with energy density $\rho_{A}$ and pressure
$ p_{A} $. The total energy density is $ \rho_{tot} = \rho + \rho_{A} $ . The conformal anomaly, also known as the trace anomaly, can be given by the trace of the fluid stress-energy tensor \begin{equation}
T_{A} = - \rho_{A} + 3 p_{A}
\end{equation} 
The conformal anomaly corrected pressure and energy densities still obey the continuity equation (2.10) and using that, we can write \begin{equation}
T_{A} = - 4 \rho_{A} - \frac{\dot{\rho_{A}}}{H} \left( \frac{1}{2 \rho_{A}} - 1 \right) 
\end{equation}  
The conformal anomaly corrected pressure and energy densities still obey the continuity equation and using that, we can write \cite{Nojiri:2005sx} \begin{equation}
T_{A} = - 4 \rho_{A} - \frac{\dot{\rho_{A}}}{H}
\end{equation}
One can then express $\rho_{A}$ as an integral in terms of $T_{A} $ as \begin{equation}
\rho_{A} = - \frac{1}{a^4} \int a^4 H T_{A} dt
\end{equation} 
Furthermore $T_{A} $ can be expressed in terms of the Hubble parameter as \begin{equation}
T_{A} = -12 b \dot{H}^{2} + 24 b^{\prime} (-\dot{H}^2 + H^{2} \dot{H} + H^{4} ) - (4b + 6 b^{\prime \prime} ) (H^{(3)} + 7 H \ddot{H} + 4 \dot{H}^2 + 12 H^{2} \dot{H})
\end{equation}
And using this, one can have an expression for $ \rho_{A} $ taking into account conformal anomaly effects near the singularity 
\begin{multline} \label{CB3}
\rho_A= -\frac{1}{a^4} \int d t\ a^4 H T_A \\ = -\frac{1}{a^4} \int d t a^4 H \Big[-12b \dot{H}^2
+ 24b' (-\dot{H}^2 + H^2 \dot{H} + H^4)  - (4b + 6b'')\left(\dddot{H}
+ 7 H \ddot{H} + 4\dot{H}^2 +
12 H^2 \dot{H} \right) \Big]
\end{multline}
While the quantum corrected Friedmann equation \footnote{Note that to maintain consistency with the notation used in \cite{Nojiri:2005sx} we are considering the Friedmann equation to be of the form $$
	H^2=\frac{\kappa^2}{3}(\rho+\rho_m)\,.$$ } is
\begin{equation}
\label{EOSq1}
\frac{3}{\kappa^2}H^2=\rho + \rho_A\,.
\end{equation}
Since the curvature is expected to be large near the time of the singularity, one can be warranted to think that $(3/\kappa^2)H^2\ll \left| \rho_A\right|$.
Then  $\rho\sim - \rho_A$ from \eqref{CB3},
which gives
\begin{equation}
\label{EOSq2}
\dot{\rho}+4H\rho =  H\Big[-12b \dot{H}^2
+ 24b' (-\dot{H}^2 + H^2 \dot{H} + H^4) - (4b + 6b'')\left(\dddot{H}
+ 7 H \ddot{H} + 4\dot{H}^2 +
12 H^2 \dot{H} \right) \Big] 
\end{equation}
Finally then, the continuity equation $\dot{\rho}+
3H\left(\rho + p\right)=0$ for $p=-\rho - f(\rho)$,
gives
\begin{equation}
\label{EOSq3}
H=\frac{\dot{\rho}}{3 f(\rho)}\,.
\end{equation}
Now we can appreciate the implications of these effects on both strong and weak singularities, where firstly we consider the big rip. The first attempt to address the issue of big rip with conformal anomalies was done in \cite{Nojiri:2003jn,Nojiri:2003vn} For this, we consider the model given by \begin{equation} \label{model1}
f(\rho)\sim B\rho^{\beta}
\end{equation} 
with $1/2<\beta<1$ when $\rho$ is large and in this case there exists the Big 
Rip singularity, as we had discussed previously too. in Sec.~IV. We note that the classical evolution is
characterized by $\rho \propto (t_s-t)^{\frac{2}{1-2\beta}}$
and $H \propto (t_s-t)^{\frac{1}{1-2\beta}}$,
both of which exhibit divergence for $\beta>1/2$. When quantum corrections are taken into account,
it is natural to assume that near the singularity $\rho$ behaves as
\begin{equation}
\label{EOSq4}
\rho = \rho_0\left(t_s - t\right)^{\tilde\gamma}
\end{equation}
As $\rho$ may diverge at $t=t_s$, we consider negative values of
$\tilde\gamma$. Since $\tilde\gamma\left(1 - \beta\right)<0$ in this case,
we might expect that \eqref{EOSq2} would give the following approximate 
relation around $t=t_s$:
\begin{equation}
\label{EOSq10}
\rho \sim 6b'H^4
\end{equation}
The term on the r.h.s. grows as
$H^4 \propto (t_s-t)^{-4+4\tilde\gamma (1-\beta)}$, but this does not give 
a
consistent result, since $\rho$ becomes negative for $b'<0$. This tells that our assumptions should be wrong and $\rho$ does not become
infinite. If $\rho$ has an extremum, \eqref{EOSq3} tells that 
$H$ vanishes there since $\dot \rho=0$. Furthermore, the authors in \cite{Nojiri:2005sx} showed numerically that in this scenario the Hubble rate approaches zero in finite time, thus coming to the conclusion that conformal anomaly effects can alleviate the Big rip in this case. 
\\
\\
Let us again consider the model in \eqref{model1} but now for the range $\beta>1$, in which case we see that a type III singularity develops with $\rho \propto \left(t_s - t\right)^{\frac{2}{1-2\beta}}$. Again, we consider that near the singularity $\rho$ behaves as \eqref{EOSq4}. Using \eqref{EOSq3}, one finds
\begin{equation}
\label{EOSq4b}
H =- \frac{\tilde\gamma \rho_0^{1-\beta}}{3B}
\left(t_s - t\right)^{-1 + \tilde\gamma\left(1 - \beta\right)}
\end{equation}
Since we are considering the case $\beta>1$ and $\tilde\gamma<0$,
we have that $\tilde\gamma\left(1 - \beta\right)>0$.
By picking up most singular term in the r.h.s of \eqref{EOSq2}, it 
follows
\begin{equation}
\label{EOSq6}
\dot{\rho} \sim -6\left(\frac{2}{3}b + b''\right)H \dddot{H}
\end{equation}
Then substituting \eqref{EOSq4} and \eqref{EOSq4b} for
\eqref{EOSq6}, we obtain
\begin{equation}
\label{gam}
\tilde\gamma=\frac{4}{1-2\beta}
\end{equation}
This means that $\rho$ and $H$ evolve as
\begin{equation}
\label{rhoHsolu}
\rho \propto (t_s-t)^{\frac{4}{1-2\beta}}\,,~~~~
H \propto (t_s-t)^{\frac{3-2\beta}{1-2\beta}}\,,
\end{equation}
around $t=t_s$. Numerically solving the background equations shows that
in the presence of quantum corrections one has
$H \propto (t_s-t)^{1/3}$ around $t=t_s$, which means that
$H$ approaches zero. Meanwhile in the absence of quantum
corrections we have $H \propto (t_s-t)^{-1/3}$, thereby showing
the divergence of $H$ at $t=t_s$. From \eqref{EOSq4b} we obtain
\begin{equation}
\label{EOSq9}
a\sim a_0\exp\left[
\frac{\rho_0^{1-\beta}}{3B\left(1-\beta\right)}\left(t_s - t\right)
^{ \tilde\gamma\left(1 - \beta\right)}
\right]
\end{equation}
where $a_0$ is a constant.
Comparing the classical case [$\tilde{\gamma}=2/(1-2\beta)$]
with the quantum corrected one [$\tilde\gamma = 4/(1-2\beta)$],
we find that the power of $\left(t_s - t\right)$ is larger in the
presence of quantum corrections. Then the scale factor approaches a constant $a_0$ more rapidly if we account for the quantum effect, implying that the
spacetime tends to be smooth, although the divergence of $\rho$
is stronger. Thus quantum effects moderate the classical singularity. 
\\
\\
But conformal anomaly effects may not always be of a huge help in order to alleviate singularities, take for example the case of an asymptotically safe cosmology which was considered in \cite{Trivedi:2022svr}. The capacity to build gravitational RG flow approximations outside of perturbation theory is necessary for conceptually testing asymptotic safety. a very strong framework for doing these calculations
is the functional renormalization group equation (FRGE) for the gravitational
effective average action $ \Gamma_{k} $ \begin{equation}
\partial_{k} \Gamma_{k} [g,\overline{g}] = \frac{1}{2} Tr \left[ (\Gamma_{k}^{(2)} + \mathcal{R}_{k} )^{-1} \partial_{k} \mathcal{R}_{k} \right]
\end{equation} 
The construction of the FRGE uses the background field formalism, where the
metric $ g_{\mu \nu} $ is split into a fixed background $ \overline{g}_{\mu \nu} $ and fluctuations $ h_{\mu \nu} $ ( see \cite{Bonanno:2017pkg} for a more detailed on asmyptotitcally safe cosmologies ). The authors in \cite{Trivedi:2022svr} considered the simplest approximation of the gravitational RG flow, which could be obtained from projecting the FRGE onto the Einstein-Hilbert action approximating $ \Gamma_{k}$ by \cite{Bonanno:2017pkg} \begin{equation}
\Gamma_{k} = \frac{1}{16 \pi G_{k}} \int d^4 x \sqrt{-g} \left[-R + 2 \Lambda_{k}\right] + \text{gauge-fixing and ghost terms}
\end{equation}
where R, $ \Lambda_{k} $ and $ G_{k} $ are the Ricci Scalar, the running cosmological constant and the running Newton's gravitational constant. The scale-dependence of these couplings can be written in terms of their dimensionless counterparts as \begin{equation}
\Lambda_{k} = k^2 \lambda_{*}
\end{equation}
\begin{equation}
G_{k} = g_{*}/k^{2}
\end{equation}
where $g_{*} = 0.707$ and $ \lambda_{*} = 0.193$ . Considering a background FLRW metric and a perfect fluid for the stress energy tensor $T_{\mu}^{\nu} = \text{diag} [-\rho,p,p,p] $ , one can get the Friedmann equation and the continuity equation in this scenario to be \begin{equation}
H^{2} = \frac{8 \pi G_{k}}{3} + \frac{\Lambda_{k}}{3}
\end{equation} 
\begin{equation} \label{asympcont}
\dot{\rho} + 3 H(\rho + p) = - \frac{\dot{\Lambda_{k}} + 8 \pi \dot{G_{k}}}{8 \pi G}
\end{equation}
Where the continuity equation comes about from the Bianchi identity which is satisfied by Einstein's equations $ D^{\mu} [\lambda(t) g_{\mu \nu} - 8 \pi G(t) T_{\mu \nu} ] = 0 $, which has the usual meaning that the divergence $ D^{\mu}$ of the Einstein tensor vanishes . The extra terms of the right hand side in \eqref{asympcont} can be interpreted as an illustration of the energy transfer between gravitational degrees of freedom and matter. Using this new continuity equation, we can write the conformal anomaly term in this case as \begin{equation} \label{tacont}
T_{A} = - 4 \rho_{A} - \frac{\dot{\rho_{A}}}{H} \left( \frac{1}{2 \rho_{A}} - 1 \right) 
\end{equation}
We note that in the conventional cosmology, one could represent the conformal anomaly corrections to the pressure in the form of an integral but it is clear that this could not be the case for the asymptotically safe cosmology as But obtaining a corresponding integral for $\rho_{A}$ for the equation \eqref{tacont}  is not possible in the same way . Hence its not feasible to address a possible removal of Type I- Type III singularities using conformal anomaly effects in this asymptotically safe cosmology.
\\
\\
\subsection{Varying constants approach}
Cosmologies with varying physical constants, like speed of light or the gravitational constant\cite{barrow1999cosmologies} have been shown to regularize cosmological singularities in certain scenarios \cite{dkabrowski2016new,Dabrowski:2012eb,Leszczynska:2014xba,Salzano:2016pny}. Here we shall discuss briefly about the fundamentals of such theories and how they can be helpful in alleviating both strong and weak cosmological singularities. Examining the generalized Einstein-Friedmann equations within the context of the theories involving the varying speed of light $c(t)$ (VSL) and varying gravitational constant $G(t)$ (VG) as presented by Barrow in their work \cite{barrow1999cosmologies}, one can deduce the following expressions for mass density $\varrho(t)$ and pressure $p(t)$:

\begin{equation}
\label{rho} \varrho(t) = \frac{3}{8\pi G(t)}
\left(\frac{\dot{a}^2}{a^2} + \frac{kc^2(t)}{a^2}
\right)
\end{equation}

\begin{equation} \label{p1}
p(t) = - \frac{c^2(t)}{8\pi G(t)} \left(2 \frac{\ddot{a}}{a} + \frac{\dot{a}^2}{a^2} + \frac{kc^2(t)}{a^2} \right)     
\end{equation}

These equations highlight the influence of varying $c$ and $G$ on mass density and pressure. For instance, if $\dot{a}$ approaches infinity while $G(t)$ increases more rapidly than $\dot{a}$, the singularity in $p(t)$ can be eliminated. In the case of flat models, a direct relationship between pressure $p$ and mass/energy density $\rho/\varepsilon$ can be established, albeit with a time-dependent equation of state parameter, expressed as:

\begin{equation}\label{pete}
p(t) = w(t) \varepsilon(t) = w(t) \varrho(t) c^2(t)    
\end{equation}

Here, the parameter $w(t)$ is defined as $w(t) = \frac{1}{3} \left[2q(t) - 1 \right]$, with $q(t) = - \ddot{a}a/\dot{a}^2$ being the dimensionless deceleration parameter. Notably, the variation of the speed of light $c(t)$ brings about a key distinction between mass density $\varrho$ and energy density $\varepsilon = \varrho c^2$, impacting the Einstein mass-energy relationship $E=mc^2$, which is transformed here into the mass density-pressure formula $p = \varrho c^2$ after division by volume. The variability of physical constants can be explored through the scale factor, allowing for the examination of scenarios like Big Bang, Big Rip, Sudden Future, Finite Scale Factor, and $w$-singularities, as expressed by the scale factor equation:

\begin{equation}
\label{newscalef} a(t) = a_s \left( \frac{t}{t_s} \right)^m \exp{\left( 1 - \frac{t}{t_s} \right)^n}
\end{equation}

The constants $t_s, a_s, m, n$ are determined accordingly \cite{Dabrowski:2012eb}. This approach illustrates how the varying constant concept aids in regularizing singularities. By inspecting Equations \eqref{rho} and \eqref{p1}, it becomes evident that a time-dependent gravitational constant variation of the form $G(t) \propto \frac{1}{t^2}$ eliminates a Type 0 Big Bang singularity in Friedmann cosmology, addressing both $p$ and $\varrho$ singularities. In Dirac's scenario \cite{dirac1937cosmological}, where $G(t) \propto 1/t$, only the $\varrho$ singularity is removed. Moreover, the time dependence of $G=1/t^2$ is less constrained by geophysical limitations on Earth's temperature \cite{teller1948change}.

Another proposal suggests that if the scale factor \eqref{newscalef} doesn't tend to zero as $t \to 0$, it could be rescaled by a "regularizing" factor $a_{rg} = (1+ 1/t^m)$ ($m \geq 0$), resulting in:

\begin{equation}
a_{sm} = \left( 1 + \frac{1}{t^m} \right) \left(\frac{t}{t_s} \right)^m = \left(\frac{t}{t_s} \right)^m + \frac{1}{t^m_s}    
\end{equation}

Consequently, a varying constant approach (in this case, related to the gravitational constant) can effectively eliminate a strong singularity, such as the Big Bang singularity. A scenario where the varying speed of light contributes to singularity regularization begins by considering a form for the ansatz of $c(t)$. One common assumption regarding the speed of light's variation is that it follows the evolution of the scale factor \cite{barrow1999cosmologies}:

\begin{equation}\label{cc0}
c(t) = c_0 a^s(t)
\end{equation}

With $c_0$ and $s$ as constants, the field equations \eqref{rho} and \eqref{p1} can be expressed as:

\begin{equation}\label{rhof}
\varrho(t) = \frac{3}{8\pi G(t)} \left(\frac{\dot{a}^2}{a^2} + kc_0^2 a^{2(s-1)} \right)
\end{equation}

\begin{equation}\label{pf}
p(t) = - \frac{c_0^2 a^{2s}}{8\pi G(t)} \left(2 \frac{\ddot{a}}{a} + \frac{\dot{a}^2}{a^2} + kc_0^2a^{2(s-1)} \right)
\end{equation}

In the presence of the time dependence of $c(t)$ as given by \eqref{cc0}, and for the choice of $a(t) = t^m$, it is possible to eliminate a pressure singularity (Type II) if certain conditions are met: $s > 1/m$ for $k=0$, $m>0$, and $s > 1/2$ or $m<0$, $s < 1/2$ for $k\neq 0$.
\\
\\
\subsection{Modified gravity effects/Quantum gravitational cosmologies}
In recent times, there has been a wide interest in dark energy models based in exotic non general relativistic regimes particularly because such theories display properties which are not evident in conventional cosmological models. For example,a lot of works have considered the possibility of viable scalar field based dark energy regimes in quantum gravity corrected cosmologies like the RS-II Braneworld and Loop Quantum Cosmology \cite{Sahni:2002dx,Sami:2004xk,Tretyakov:2005en,Chen:2008ca,Fu:2008gh}. There has been substantial work on new dark energy models based on thermodynamic modifications like modified area-entropy relations \cite{Tavayef:2018xwx,Radicella:2011qpl,Bamba:2009id,Younas:2018kmy,Jawad:2016tne,Nojiri:2019skr} or even more exotic possibilities like generalized uncertainty principles  \cite{Ghosh:2011ft,Rashki:2019mde,Paliathanasis:2021egx} or non-canonical approaches like DBI etc. as well \cite{Calcagni:2006ge,Gumjudpai:2009uy,Chiba:2009nh,Ahn:2009xd,Li:2016grl,mandal2021dynamical,Kar:2021gbz}.This vast dark energy literature has prompted the study of cosmological singularities in a wide range of cosmological backgrounds as well, as there have been multiple works which have discussed Type I-IV singularities in various cosmologies \cite{Shtanov:2002ek,Bamba:2012ka,Bamba:2010wfw,Nojiri:2008fk,odintsov2018dynamical,Odintsov:2018awm,trivedi2022finite,Bombacigno:2021bpk,Nojiri:2006gh,Fernandez-Jambrina:2021foi,Chimento:2015gga,Chimento:2015gum,Cataldo:2017nck,Nojiri:2005sr,Nojiri:2005sx,Trivedi:2022svr,trivedi2022type,trivedi2022type1}. In this vast array of works, one has seen quite a few examples in which cosmologies affected by the effects of these modified gravity theories or quantum gravitational paradigms ( like the Braneworld or LQC for example ) have alleviated certain singularities. Here we would like to consider an example of how such effects can help in alleviating type V singularities, as we have not discussed ways to remove or moderate this till now. 
\\
\\
We would like to consider the treatment in \cite{trivedi2022type1} for our example here. We would like to again consider a model with an inhomogeneous EOS of the form $ p = - \rho - f(\rho) $. It was shown in \cite{Dabrowski:2012eb} that for the simplified case of the scale factor \eqref{wansatz} with $m=0$, one can get w-singularities for $ n > 0 $ and $ n \neq 1 $. The scale factor for the case $ m = 0 $ takes the form \begin{equation}
a(t) = a_{s} \exp \left(1 - \frac{t}{t_{s}}\right)^{n}
\end{equation} and we will be using this form of the scale factor for this example. The modified gravity theory we would be interested in is an f(R) gravity model with the action \cite{Carroll2004} \begin{equation} \label{fraction}
S = \frac{m_{p}^2}{2} \int d^4 x \sqrt{-g} \left(R - \frac{\alpha^2}{R}\right) + \int d^4 x \sqrt{-g} \mathcal{L}_{m}
\end{equation}
where $\alpha$ is a constant which has the units of mass, $\mathcal{L}_{m} $ is the Lagrangian density for matter and $m_{p}$ is the reduced Planck's constant. The field equation for this action is 
\begin{equation} \label{actionfr}
\left(1 + \frac{\alpha^2}{R^2}\right) R_{\mu \nu} - \frac{1}{2} \left(1 - \frac{\alpha^2}{R^2}\right) R g_{\mu \nu} + \alpha^2 \left[g_{\mu \nu} \nabla_{a} \nabla^{a} - \nabla (_{\mu} \nabla_{\nu})  \right] R^{-2} =  \frac{T_{\mu \nu}^{M}}{m_{p}^2}
\end{equation}
The Friedmann equation in this case can take the form \begin{equation} \label{frfriedmann}
\frac{6 H^2 - \frac{\alpha}{2}}{11/8 - \frac{8 H^2}{4 \alpha}} = \frac{\rho}{3}
\end{equation} where $\rho$ is the total energy density. This $F(R)$ gravity regime was used to explain late time cosmic acceleration as an alternative to dark energy in \cite{Carroll2004}. The use of f(R) gravity regimes for avoiding cosmological singularities by adding an $R^2$ term was considered in detail in \cite{abdalla2005consistent}, with the same scenario later being extended in \cite{Bamba:2008ut} and  \cite{Nojiri:2008fk}. 
Moreover, based on properties of $R^2$ term, non-singular modified gravity were proposed in \cite{elizalde2011nonsingular}. The action \eqref{actionfr} prompts one towards the notion that very tiny corrections to the usual Einstein Hilbert in the form of $R^{n}$ with $n<0$ can produce cosmic acceleration. As corrections of the form $R^n$ with $n>0$ can lead to inflation in the early universe \cite{Starobinsky:1980te}, the authors in \cite{Carroll2004} proposed a purely gravitational paradigm through \eqref{fraction} to explain both the early and late time accelerations of the universe. 
\\
Now we consider $f(\rho) = \rho^{\alpha} $ and firstly see the status quo of w-singularities for such a model in the standard cosmology given by ( written here in natural units just for simplicity) $$H^2 = \frac{\rho }{3}$$. We can write the w-parameter for this cosmology using as \begin{equation}
w = -3^{\alpha-1} \left(\frac{n^2 \left(1-\frac{t}{t_{s}}\right)^{2 (n-1)}}{t_{s}^2}\right)^{\alpha-1}-1
\end{equation}
From this we can make the following observations : \begin{itemize}
	\item For n = 1, no w- singularities occur as is the case in the usual scenario with conventional equation of state. 
	\item For $\alpha < 0 $ , w-singularities occur for all values of positive values of n besides unity but w-singularities do not occur for any negative values of n
	\item For $\alpha > 0 $ we see a very interesting behaviour. In this case, completely in contrast to what happens in the usual case, no w-singularities occur for positive values of n ($n>0$ ) but they occur only when n has negative values ($n<0$). Hence, here we see the first sign of departure in the occurrence conditions of w-singularities when one considers inhomogeneous equations of state.  
\end{itemize}
So we see here that incorporating an inhomogeneous EOS can be of use in moderating w-singularities, but this still does not remove them per say as it only changes the conditions under which they occur with regards to what happens in the conventional cosmology. Now the w-parameter for \eqref{frfriedmann} case is given by \begin{equation}
w = -12^{\alpha-1} \left(\frac{\eta \left(12 n^2 \left(1-\frac{t}{t_{s}}\right)^{2 n}-\eta (t_{s}-t)^2\right)}{11 \eta (t_{s}-t)^2-18 n^2 \left(1-\frac{t}{t_{s}}\right)^{2 n}}\right)^{\alpha-1}-1
\end{equation} 
For the w-parameter as expressed above, we have the following observations : \begin{itemize}
	\item For n = 1, contrary to the other cases we have considered till now, one can have a w-singularity but that is possible only in the extreme case that $\alpha \to \infty$ which is not pretty realistic to expect but in principle singularities can appear in this case. 
	\item The most interesting thing that comes out when one considers this scenario is that w-singularities do not occur for any value of n and $\alpha$ ! For both positive and negative values of $\alpha$ and n, the w-parameter remains regular and does not diverges.
\end{itemize}
And so we see that just by incorporating the effects of a modified gravity theory, in this case a particular form of $f(R)$ gravity, one can alleviate singularities too. Furthermore, $f(R)$ gravity theories which been of great use in alleviating various other singularities too but we have discussed about other singularities quite extensively before and so it seems appropriate to discuss this example to illustrate how type V singularities could be moderated too.
\\
\\
\section{Dynamical systems approach and the Goriely-Hyde method}
While it is seems pretty natural to study singularities and their avoidance methods in various cosmological settings like we have discussed so far,often it is very difficult to classify and study the cosmological singularities which may occur in extremely non-conventional cosmologies which are motivated by quantum gravitational/phenomenological considerations (for example, see the classification of singularities in asymptotically safe cosmology \cite{Trivedi:2022svr} ) and often it may not even be possible to do so in an orthodox fashion. Hence it becomes essential to look for non-conventional ways to find out cosmological singularities in exotic cosmologies and in this regard, a particular dynamical systems method can be of huge help. From a dynamical standpoint, one of the most intriguing aspects of studying various dynamical systems lies in understanding their singularity structure, which becomes particularly relevant when these systems describe physically significant phenomena. While numerous approaches have been proposed to explore the singularity structure of autonomous dynamical systems, one particularly interesting method is the Goriely-Hyde procedure \cite{goriely2000necessary}. As cosmology presents a multitude of captivating dynamical systems \cite{bahamonde2018dynamical}, the investigation of singularity structure in such systems has gained considerable attention, with the Goriely-Hyde method proving particularly useful for cosmological explorations \cite{barrow2004more,Cotsakis:2006pi,cotsakis2007asymptotics,antoniadis2010brane,antoniadis2013brane,antoniadis2014enveloping}. This method has previously been applied to study finite and non-finite time singularities in certain classes of quintessence models as well \cite{odintsov2018dynamical,odintsov2019finite,trivedi2022finite}.The Goriely-Hyde method provides an elegant approach to determining the presence of singularities in dynamical systems and the procedure can be outlined as follows:

\begin{itemize}
	\item We begin by considering a dynamical system described by $n$ differential equations of the form:
	\begin{equation}
	\dot{x}_{i} = f_{i}(x),
	\end{equation}
	where $i = 1, 2, ..., n$, and the overdot represents differentiation with respect to time $t$, which in the case of quintessence models can be better represented by the number of e-foldings $N$. We identify the parts of the equation $f_{i}$ that become significant as the system approaches the singularity. These significant parts are referred to as "dominant parts" \cite{goriely2000necessary}. Each dominant part constitutes a mathematically consistent truncation of the system, denoted as $\hat{f}_{i}$. The system can then be written as:
	\begin{equation}
	\dot{x}_{i} = \hat{f}_{i}(x).
	\end{equation}
	
	\item Without loss of generality, the variables $x_{i}$ near the singularity can be expressed as:
	\begin{equation}
	x_{i} = a_{i} \tau^{p_{i}},
	\end{equation}
	where $\tau = t - t_{c}$, and $t_{c}$ is an integration constant. Substituting equation (4) into equation (3) and equating the exponents, we can determine the values of $p_{i}$ for different $i$, which form the vector $\mathbf{p} = (p_{1}, p_{2}, ..., p_{n})$. Similarly, we calculate the values of $a_{i}$ to form the vector $\vec{a} = (a_{1}, a_{2}, ..., a_{n})$. It is important to note that if $\vec{a}$ contains only real entries, it corresponds to finite-time singularities. Conversely, if $\vec{a}$ contains at least one complex entry, it may lead to non-finite-time singularities. Each set $(a_{i}, p_{i})$ is known as a dominant balance of the system.
	
	\item Next, we calculate the Kovalevskaya matrix given by:
	\begin{equation}
	R = \begin{pmatrix}
	\frac{\partial f_{1}}{\partial x_{1}} & \frac{\partial f_{1}}{\partial x_{2}} & . & . & \frac{\partial f_{1}}{\partial x_{n}}\\
	\frac{\partial f_{2}}{\partial x_{1}} & \frac{\partial f_{2}}{\partial x_{2}} & . & . & \frac{\partial f_{2}}{\partial x_{n}}\\
	. & . & . & . & . \\
	. & . & . & . & . \\
	\frac{\partial f_{n}}{\partial x_{1}} & \frac{\partial f_{n}}{\partial x_{2}} & . & . & \frac{\partial f_{n}}{\partial x_{n}}\\
	\end{pmatrix} -  \begin{pmatrix}
	p_{1} & 0 & . & . & 0 \\
	0 & p_{2} & . &
	
	. & 0 \\
	. & . & . & . & . \\
	. & . & . & . & . \\
	0 & 0 & . & . & p_{n} \\
	\end{pmatrix}.
	\end{equation}
	
	After obtaining the Kovalevskaya matrix, we evaluate it for different dominant balances and determine the eigenvalues. If the eigenvalues are of the form $(-1, r_{2}, r_{3}, ..., r_{n})$, with $r_{2}, r_{3}, ... > 0$, then the singularity is considered general and will occur regardless of the initial conditions of the system. Conversely, if any of the eigenvalues $r_{2}, r_{3}, ...$ are negative, the singularity is considered local and will only occur for certain sets of initial conditions.
	\item Without loss of generality, the variables $x_{i}$ near the singularity can be expressed as:
	\begin{equation}
	x_{i} = a_{i} \tau^{p_{i}},
	\end{equation}
	where $\tau = t - t_{c}$, and $t_{c}$ is an integration constant. Substituting equation (4) into equation (3) and equating the exponents, we can determine the values of $p_{i}$ for different $i$, which form the vector $\mathbf{p} = (p_{1}, p_{2}, ..., p_{n})$. Similarly, we calculate the values of $a_{i}$ to form the vector $\vec{a} = (a_{1}, a_{2}, ..., a_{n})$. It is important to note that if $\vec{a}$ contains only real entries, it corresponds to finite-time singularities. Conversely, if $\vec{a}$ contains at least one complex entry, it may lead to non-finite-time singularities. Each set $(a_{i}, p_{i})$ is known as a dominant balance of the system.
	
	\item Next, we calculate the Kovalevskaya matrix given by:
	\begin{equation}
	R = \begin{pmatrix}
	\frac{\partial f_{1}}{\partial x_{1}} & \frac{\partial f_{1}}{\partial x_{2}} & . & . & \frac{\partial f_{1}}{\partial x_{n}}\\
	\frac{\partial f_{2}}{\partial x_{1}} & \frac{\partial f_{2}}{\partial x_{2}} & . & . & \frac{\partial f_{2}}{\partial x_{n}}\\
	. & . & . & . & . \\
	. & . & . & . & . \\
	\frac{\partial f_{n}}{\partial x_{1}} & \frac{\partial f_{n}}{\partial x_{2}} & . & . & \frac{\partial f_{n}}{\partial x_{n}}\\
	\end{pmatrix} -  \begin{pmatrix}
	p_{1} & 0 & . & . & 0 \\
	0 & p_{2} & . & . & 0 \\
	. & . & . & . & . \\
	. & . & . & . & . \\
	0 & 0 & . & . & p_{n} \\
	\end{pmatrix}.
	\end{equation}
	
	After obtaining the Kovalevskaya matrix, we evaluate it for different dominant balances and determine the eigenvalues. If the eigenvalues are of the form $(-1, r_{2}, r_{3}, ..., r_{n})$, with $r_{2}, r_{3}, ... > 0$, then the singularity is considered general and will occur regardless of the initial conditions of the system. Conversely, if any of the eigenvalues $r_{2}, r_{3}, ...$ are negative, the singularity is considered local and will only occur for certain sets of initial conditions.
\end{itemize}
After applying the method, one can then classify singularities using well motivated ansatz' for the scale factor or the Hubble parameter.The most general form of the Hubble parameter for investigating singularities within the aforementioned classified types is expressed as \cite{odintsov2019finite}:

\begin{equation} \label{a11}
H(t) = f_{1}(t) + f_{2}(t)(t - t_{s})^{\alpha}
\end{equation}

Here, $f_{1}(t)$ and $f_{2}(t)$ are assumed to be nonzero regular functions at the time of the singularity, and similar conditions apply to their derivatives up to the second-order. Additionally, $\alpha$ is a real number. It is not mandatory for the Hubble parameter (34) to be a solution to the field equations; however, we will consider this case and explore the implications of this assumption on the singularity structure based on our dynamic analysis. First, we observe that none of the variables $x$, $y$, or $z$ as defined in (10) can ever become singular for any cosmic time value. The singularities that can occur considering the Hubble parameter as defined in (34) are as follows:

\begin{itemize}
	\item For $\alpha < -1$, a big rip singularity occurs.
	\item For $-1 < \alpha < 0$, a Type III singularity occurs.
	\item For $0 < \alpha < 1$, a Type II singularity occurs.
	\item For $\alpha > 1$, a Type IV singularity occurs.
\end{itemize}

Another ansatz useful for classifying singularities was introduced in \cite{Odintsov:2022eqm} whereby the scale factor was written as \begin{equation} \label{a12}
a(t) = g(t) (t-t_{s})^{\alpha} + f(t)
\end{equation}
where g(t) and f(t) and all their higher order derivatives with respect to the cosmic time are smooth functions of the cosmic time. For this ansatz, according to the values of the exponent $\alpha$ one can have the following singularities \begin{itemize}
	\item For $\alpha < 0 $, a type I singularity occurs 
	\item For $0 < \alpha < 1$, a type III singularity develops
	\item For $a < \alpha < 2$, a type II singularity occurs 
	\item For $\alpha > 2$, a type IV singularity occurs 
\end{itemize}
Again, it is not mandatory that the scale factor in \eqref{a12} will necessarily be a solution to the field equations but we would like to consider this and \eqref{a11} in order to get a well-motivated feel for the type of cosmological singularities we can deal with in the various models we have discussed so far. 
As an example for this method, let's consider singularities in an RS-II braneworld cosmology where dark energy can be described by a scalar field paradigm, where we shall follow the treatment of \cite{trivedi2022finite}. The action for inclusive of both the scalar and the background fluid term can be written as \begin{multline}
S  = S_{RS} + S_{B} + S_{\phi} = \int d^5 x \sqrt{-g^{(5)}} \left( \Lambda^{(5)}  + 2 R^{(5)}   \right) + \\ \int d^4 x \sqrt{-g} \left(\sigma -\frac{1}{2} \mu(\phi) (\nabla \phi)^2 - V(\phi)  + \mathcal{L}_{B}  \right)
\end{multline}
where $R^{(5)} $, $ g^{(5)}_{\mu \nu} $ and $ \Lambda^{(5)} $ are the bulk Ricci Scalar, metric and the cosmological constant respectively with $\sigma$ being the brane tension on the 3-brane, $g_{\mu \nu} $ being the 3-brane metric and $\mu(\phi) $ being a scalar coupling function. Note that here we are working in Planck units with $(m_{p}^{(5)})^2 = 1 $ with $ m_{p}^{(5)} $ being the 5-dimensional Planck mass. Assuming that the brane metric has the usual FLRW form, we get the Friedmann equation to be \cite{Maartens:2010ar} \begin{equation}
H^{2}  = \rho \left(1 + \frac{\rho}{2 \sigma}\right)
\end{equation} 
where $ \rho = \rho_{\phi} + \rho_{B} $ is the total cosmological energy density taking into account contributions from both the scalar field and the background fluid term and the Bulk cosmological constant has been set to zero for simplicity. One can similarly find \begin{equation}
2 \dot{H} = -\left(1 + \frac{\rho}{\sigma}\right) \left(\mu(\phi) \dot{\phi}^2 + \rho_{B} \right)
\end{equation}
And the equation motion of the scalar is given by \begin{equation}
\mu(\phi) \ddot{\phi} + \frac{1}{2} \frac{d \mu}{d \phi} \dot{\phi}^2  + 3 H \mu(\phi) \dot{\phi} + \frac{dV}{d\phi} = 0 
\end{equation}
Finally, using the following variables introduced in \cite{Gonzalez:2008wa} \begin{equation}
x = \frac{\dot{\phi}}{\sqrt{6}H} \qquad y = \frac{\sqrt{V}}{\sqrt{3} H} \qquad z = \frac{\rho}{3 H^2}
\end{equation}
Choosing the background fluid to be of the form of pressurelees dark matter, in a way that $w_{B} = 0 $ , we get the dynamical system for this model to be \begin{equation}
x^{\prime} = - \sqrt{\frac{3}{2 \mu }} \lambda y^2 -3x + \frac{3x}{2} \left(z + x^2 -y^2 \right) \left(\frac{2}{z} - 1 \right)
\end{equation}
\begin{equation}
y^{\prime} = \sqrt{\frac{3}{2 \mu }} \lambda xy + \frac{3y}{2} \left(z + x^2 -y^2 \right) \left(\frac{2}{z} - 1 \right)
\end{equation}
\begin{equation}
z^{\prime} = 3 (1-z) (z + x^2 - y^2 )
\end{equation}
where the primes denote differentiation with respect to the e-folding number N and $\lambda = \frac{V^{\prime}}{V}$. We can finally start with the analysis as we have proper autonomous dynamical system, with the first truncation that we consider being \begin{equation}
\hat{f} = \begin{pmatrix}
-k \lambda y^2 \\
-3 y^3 z^{-1} \\
3 x^2
\end{pmatrix}
\end{equation}
where $ k = \sqrt{\frac{3}{2 \mu}} $. Using the ansatz of the Goriely-Hyde method, we get $ \mathbf{p} = (-1,-1,-1) $ and using these, we get \begin{equation}
\begin{aligned}
a_{1} = \left(- \frac{1}{k \lambda} , \frac{i}{k \lambda} , - \frac{3}{k^2 \lambda^2} \right) \\[10pt]
a_{2} = \left( - \frac{1}{k \lambda} ,- \frac{i}{k \lambda} , - \frac{3}{k^2 \lambda^2} \right)  
\end{aligned}
\end{equation}
as both $ a_{1} $ and $a_{2}$ have complex entries, only non-finite time singularities will be possible with regards to this truncation. The Kovalevskaya matrix then takes the form \begin{equation}
R =  \begin{pmatrix}
1 & -2k\lambda y & 0 \\
0 & 1 - \frac{9 y^2}{z} & \frac{3 y^{3}}{z^2} \\
6x & 0 & 1 
\end{pmatrix}
\end{equation}
We then finally find the eigenvalues of the matrix, which are given by \begin{equation}
r = (-1,-1,2)
\end{equation}
Hence the singularities in this case will only be local singularities which will only form for a limited set of initial conditions. In \cite{trivedi2022finite}, it was worked out that there are two more possible truncations, with the balances and corresponding eigenvalues in them being \begin{equation} \label{trunc1}
\begin{aligned}
a_{1} = \left(\frac{1}{\sqrt{3}} , \frac{i}{\sqrt{3}} , \frac{1}{3} \right) \\[10pt]
a_{2} = \left(\frac{1}{\sqrt{3}} , -\frac{i}{\sqrt{3}} , \frac{1}{3} \right) \\[10pt]
a_{3} = \left(-\frac{1}{\sqrt{3}} , \frac{i}{\sqrt{3}} , \frac{1}{3} \right) \\[10pt]
a_{4} = \left(-\frac{1}{\sqrt{3}} , -\frac{i}{\sqrt{3}} , \frac{1}{3} \right)
\end{aligned}
\end{equation} with eigenvalues being \begin{equation} \label{r1}
r = (-1, \sqrt{\frac{3}{2}}, -\sqrt{\frac{3}{2}})
\end{equation}
And another truncation with the balance \begin{equation} \label{trunc2}
\begin{aligned}
a_{1} = \left(\frac{1}{\sqrt{3}} , \sqrt{\frac{2}{3}} , \frac{2}{3} \right) \\[10pt]
a_{2} = \left(\frac{1}{\sqrt{3}} , -\sqrt{\frac{2}{3}} , \frac{2}{3} \right) \\[10pt]
a_{3} = \left(-\frac{1}{\sqrt{3}} , \sqrt{\frac{2}{3}} , \frac{2}{3} \right) \\[10pt]
a_{4} = \left(-\frac{1}{\sqrt{3}} , -\sqrt{\frac{2}{3}} , \frac{2}{3} \right)
\end{aligned}
\end{equation}
with \begin{equation} \label{r2}
r = (-1,-1,1)
\end{equation}
We see from \eqref{trunc1} and \eqref{r1} that the truncations for which they belong to seem to still tell the story that only non-finite time local singularities could be possible in the system but we note from \eqref{trunc2}, that the other truncation will allow for finite time singularities albeit they would still be local as \eqref{r2} has $r_{2}=-1$. To proceed further and now classify the singularities physically, we use the ansatz for the Hubble parameter \eqref{a11} and we need to express $ \dot{\phi} $ and $ V(\phi) $ in terms of the Hubble parameter. For simplicity, we will consider that the coupling constant $ \mu = 1 $ and $ \dot{\rho_{B}} = 0 $. Making these considerations, we can write  \begin{equation}
-2 \dot{H} = \dot{\phi}^2 \left(1 + \frac{\rho}{\sigma}\right)
\end{equation}
One can then write \begin{equation}
\dot{\phi}^2 = -2 \left[ \left(\sigma +  V +  \sigma \rho_{B} \right) + \sqrt{\left( \sigma +  V +  \sigma \rho_{B} \right)^{2} - 2 \dot{H}} \right]
\end{equation} 
Furthermore, one can now write $ V(\phi)$ in terms of the dark energy equation of state \footnote{Note that here we are only considering dark energy equation of state with no background contributions, hence here we will only consider scalar field contributions} as \begin{equation}
V(\phi) = \frac{\dot{\phi}^{2}}{2} \frac{(1-w)}{(1+w)}
\end{equation}
Then we can write the potential as \begin{equation}
V = \frac{2 b (1+k) + \sqrt{(2b(1+k))^{2} - 2 \dot{H} (k^2 - 1)}}{2 (k^2 -1 )}
\end{equation}
where $k = \frac{2w}{1-w} $ and $b = \sigma (1+ \rho_{B}) $ (note that both k and b will always be positive for a positive brane tension). Notice that V is now completely in terms of the Hubble parameter (for constant values of $\sigma$, w and $\rho_{B} $) and so one can use this form of V in to find $\dot{\phi}$ in terms of the Hubble parameter as well. It is necessary to express these quantities in terms of $H(t) $ as now we can find out which type of singularities are possible in this scenario, in the view of the fact that x,y and z as described in have to remain regular. By studying the expressions for these variables, one can make out that Type I, Type III and Type IV singularities are allowed in this scenario while Type II is not. This also makes us realize that even if the cosmology is heavily motivated by quantum gravitational considerations( like the Braneworld in this scenario), it can still have quite a few cosmological singularities.
\section{Future outlooks and conclusions}
In this brief review paper, we have discussed (almost) all the prominent developments in the field of cosmological singularities which have taken place in the recent 25 years or so.  We firstly provided a detailed outlook on what space-time singularities are and discussed the various nuances regarding them, like various strength criterion etc. After that, we discussed in detail about the prominent strong and weak cosmological singularities in accordance with the classification scheme provided by Odintsov and Nojiri. We detailed under which conditions these singularities can occur in various scenarios and under which cosmological settings they were initially found in. We then saw how one can moderate or even remove these singularities using various techniques having quantum or modified gravity origins and we also discussed the Goriely-Hyde method and its usefulness in singularity works. As a whole, one general point that we could safely make is that such singularities provide a revealing arena on the interface of cosmology and quantum gravitational theories. The scales at which such events could take place lie in the horizons for testing quantum gravity ideas and with the constant increase in the precision of various observational setups, one would not be wrong to think that investigating such singularities in detail can possibly shed light on problems in both cosmology and quantum gravity.
\\
\\
\section*{Acknowledgments}
The author would like to thank Sergei Odintsov for the invitation to write this review article and for the numerous discussions with him on various aspects related to singularities. I would also like to thank Maxim Khlopov, Pankaj Joshi, Robert Scherrer, Alexander Timoshkin, Vasilis Oikonomou, Jackson Levi Said and Rafael Nunes for various discussions on singularities. I would also like to thank Parth Bhambhaniya for discussions on some particular aspects of singularities. Finally, I would also like to thank Sunny Vagnozzi for discussions on cosmology and dark energy in general which have been very helpful for this work. 
\\
\\
\section*{Appendix A }
The use of power series expansions for the scale factor and related quantities in cosmology has gathered significant pace in recent times(for a detailed overview, see \cite{Cattoen:2005dx}) and especially in the context of studies of cosmological singularities. Hence it is fitting to discuss such expansions in a bit of detail here. Generalized Frobenius series find frequent application in the expansion of solutions to differential equations around their singular points. With this characteristic in mind, we will assume that in the vicinity of the key point, the scale factor exhibits an expansion based on a generalized power series. This concept extends the familiar notions of Taylor series, meromorphic Laurent series, Frobenius series, and Liapunov expansions, as referenced in~\cite{solomon1977differential}. Furthermore, this generalized power series is more encompassing than the one employed in~\cite{Visser:2002ww}.

In the current context, if the scale factor $a(t)$ can be expressed using such a generalized power series, then the Friedmann equations dictate that both $\rho(t)$ and $p(t)$ can likewise be represented using such power series. Employing formal series reversion, it follows that the equation of state $\rho(p)$, and consequently the function $\rho(a)$, exhibit these generalized power series expansions. Conversely, when $\rho(a)$ is described by such a generalized power series, the first Friedmann equation indicates that $\dot a(t)$ possesses a power series of similar nature, which, upon integration, implies that $a(t)$ itself is characterized by such a power series.

Similarly, if the equation of state $p(\rho)$ can be expressed as a generalized power series, then integrating the conservation equation leads to the expression:

\begin{equation}
a(\rho) =   a_{0}  \exp\left\{ \frac{1}{3} \int_{\rho_{0}}^{\rho} \frac{d \bar{\rho}}{\bar{\rho} + p(\bar{\rho})}\right\}, 
\end{equation}

This equation also adopts a generalized power series representation. The potential value of expanding the conventional notion of a Frobenius series becomes evident through the analysis presented in~\cite{Visser:2002ww}.

It is important to clarify the types of entities that fall outside this category of generalized power series. First, essential singularities, effectively infinite-order poles that emerge, for instance, in functions like $\exp(-1/x)$ near $x=0$, lie beyond this classification. Secondly, certain variations on the concept of Puiseux series, specifically those containing terms like $(\ln x)^n$, $(\ln\ln x)^n$, $(\ln\ln\ln x)^n$, and so forth, also exist beyond this classification. However, there is currently no awareness of any scenarios where these exceptional cases become pertinent in a physical context.  
\\
\\
It has been shown to be reasonable to assume that in the vicinity of some cosmological singularity, happening at some time $t_{0}$ , the scale factor has a (possibly one-sided) generalized power series expansion of the form
\begin{equation}
a(t) = c_0 |t-t_{0}|^{\eta_0} + c_1  |t-t_{0}|^{\eta_1} + c_2  |t-t_{0}|^{\eta_2} 
+ c_3 |t-t_{0}|^{\eta_3} +\dots
\end{equation}
where the indicial exponents $\eta_i$ are generically real (but are often non-integer) and without loss of generality are ordered in such a way that they satisfy
\begin{equation}
\eta_0<\eta_1<\eta_2<\eta_3\dots
\end{equation}
Finally we can also without loss of generality set
\begin{equation}
c_0 > 0.
\end{equation}
There are no \emph{a priori} constraints on the signs of the other $c_i$, though by definition $c_i\neq0$.
\\
\\
From a physical point of view, this definition is really generic and can be applied to any type of cosmological milestone.
This generalized power series expansion is sufficient to encompass all the models we are aware of in the literature, and as a matter of fact, the indicial exponents $\eta_i$ will be used to classify the type of cosmological singularity we are dealing with.   
For many of the calculations in this chapter, the first term in the expansion is dominant, but even for the most subtle of the explicit calculations below  it will be sufficient to keep only the first three terms of the expansion:
\begin{equation}
a(t) = c_0 |t-t_{0}|^{\eta_0} + c_1  |t-t_{0}|^{\eta_1} + c_2  |t-t_{0}|^{\eta_2} \dots;
\qquad
\eta_0 < \eta_1<\eta_2; 
\qquad c_0 > 0.
\end{equation}
The lowest few of the indicial exponents are sufficient to determine the relationship between these cosmological milestones, the curvature singularities and even the energy. Note also that this expansion fails if the cosmological milestone is pushed into the infinite past or infinite future. Using such an expansion, one can encounter quite a few cosmological singularities and we shall list some conditions under which some prominent singularities \footnote{It is worth noting that with such a power series ansatz for the scale factor, one can also find conditions in which other exotic cosmological scenarios like bounce, emergent universe etc. can also be recovered but here we shall not list that as we are only interested in cosmological singularities. } can be found as follows : \begin{itemize}
	\item Big bang (Type 0):  If a big bang occurs at time $t_{0}$ \footnote{Similar series can be used for the Big crunch too, in which case the series takes the form $$a(t) =  c_0 (t_{0}-t)^{\eta_0} + c_1 (t_{0}-t)^{\eta_1} +\dots$$ with $t_{0}$ being the time of the big crunch.}, we define the behavior with indicial exponents ($0<\eta_0<\eta_1\dots$) when the scale factor has a generalized power series near the singularity, given by:
	
	\begin{eqnarray}
	a(t) =  c_0 (t-t_{0})^{\eta_0} + c_1 (t-t_{0})^{\eta_1} +\dots
	\end{eqnarray}
	
	The series is carefully constructed such that $a(t_{0})=0$.
	
	\item Big rip (Type 1): If a big rip occurs at time $t_{0}$, the indicial exponents of the rip ($\eta_0<\eta_1\dots$) are defined when the scale factor has a generalized power series near the rip:
	
	\begin{eqnarray}
	a(t) = c_0  |t_{0}-t|^{\eta_0} + c_1  |t_{0}-t|^{\eta_1} + \dots,
	\end{eqnarray}
	
	where $\eta_0<0$ and $c_0>0$. The series is constructed to satisfy $a(t_{0})=\infty$. The only difference from the big bang case is the \emph{sign} of the exponent $\eta_0$.
	
	\item Sudden singularity (Type II): If a sudden singularity occurs at time $t_{0}$ (past or future), the exponent is defined as $\eta_0=0$ and $\eta_1>0$, resulting in the scale factor's generalized power series near the singularity:
	
	\begin{eqnarray}
	a(t) =  c_0  + c_1 |t-t_{0}|^{\eta_1} + \dots
	\end{eqnarray}
	
	Here, $c_0>0$ and $\eta_1$ is a non-integer. The condition $a(t_{0})=c_0$ ensures finiteness, and a sufficient number of differentiations yields:
	
	\begin{equation}
	a^{(n)}(t\to t_{0}) \sim c_0 \; \eta_1 (\eta_1-1)(\eta_1-2)\dots (\eta_1-n+1) \; |t-t_{0}|^{\eta_1-n}\to\infty.
	\end{equation}
	
	The toy model by Barrow \cite{Barrow:2004xh} can be expressed as:
	
	\begin{equation}
	a(t) = c_0 \left[ (t_{0}-t)^\eta -1 \right] + \tilde c_0 (t-t_{b})^{\tilde\eta}
	\end{equation}
	
	where $t_{b}$ is the time of the big bang. This model fits into the general classification when expanded around the sudden singularity time, $t_{0}$, and into the classification of big bang singularities when expanded around the big bang time, $t_{b}$.
\end{itemize}
\section*{Appendix B}
Over the years, several alternatives to the Big rip have been found and the first one that we shall discuss in this regard is the Little rip (LR). It is characterized by a growing energy density $\rho$ over time, but this increase follows an asymptotic pattern, necessitating an infinite amount of time to approach the singularity. This situation corresponds to an equation of state parameter $w$ that falls below -1; however, it approaches -1 as time progresses towards infinity. The energy density's growth is gradual, preventing the emergence of the Big Rip singularity. The LR models depict transitional behaviors between a asymptotic de Sitter expansion and a BR evolution. In their work \cite{frampton2012models}, the authors presented an elegant method for comprehending the implications of the little rip, distinguishing it from the Big Rip, which we will explore in the following.

During the universe's expansion, the relative acceleration between two points separated by a comoving distance $l$ can be expressed as $l \ddot a/a$, where $a$ signifies the scale factor. If an observer is situated at a comoving distance $l$ from a mass $m$, they will detect an inertial force acting on the mass as follows:

\begin{equation}
\label{i1}
F_\mathrm{iner}=m l \ddot a/a = m l \left( \dot H + H^2 \right)
\end{equation} 

Let's assume that the two particles are bound by a constant force $F_0$. When the positive value of $F_\mathrm{iner}$ surpasses $F_0$, the two particles become unbound. This scenario corresponds to the phenomenon known as the "rip," which emerges due to the accelerating expansion. Equation \eqref{i1} demonstrates that a rip always occurs when either $H$ or $\dot H$ diverges (assuming $\dot H > 0$). The divergence of $H$ leads to a "big rip", while if $H$ remains finite but $\dot H$ diverges with $\dot H > 0$, it results in a Type II or "sudden future" singularity \cite{Barrow:2004xh,barrow2004more,Nojiri:2005sx}, which also causes a rip.

Nonetheless, as pointed out in \cite{Frampton:2011sp}, it's feasible for $H$ and, consequently, $F_\mathrm{iner}$, to grow boundlessly without inducing a future singularity at a finite time. This phenomenon is referred to as the little rip. Both the big rip and the little rip share the characteristic of $F_\mathrm{iner} \rightarrow \infty$; the distinction lies in the fact that for the big rip, $F_\mathrm{iner} \rightarrow \infty$ occurs at a finite time, whereas for the little rip, it occurs as $t \rightarrow \infty$. Two possible ansatz/models which have been shown to lead to little rip behaviour \cite{frampton2012models} are given by the following forms of the Hubble parameter \begin{equation}
H(t) = H_{0} \exp{\lambda t}
\end{equation} where $H_{0}$ and $\lambda$ are positive constants while another viable model which is similar to this is given by \begin{equation}
H(t) = H_{0} \exp{C \exp{\lambda t}} 
\end{equation} where $H_{0}$, $\lambda$ and $C$ are positive constants as well.  Another interesting possibility for the evolution of the universe is the so-called Pseudo-Rip \cite{frampton2012pseudo}, where the Hubble parameter, although increasing, tends to a "cosmological constant" in the remote future. That means, $H(t) \rightarrow H_\infty <\infty, t\rightarrow +\infty$, where $H_\infty$ is a constant. A possible model for this is given by the Hubble ansatz given as \begin{equation}
H(t) = H_{0} - H_{1} \exp{-\lambda t}
\end{equation} where $H_{0}$, $H_{1}$ and $\lambda$ are positive constants with $H_{0} > H_{1} $. Yet another possible alternative for the rip is a model in which the dark energy density $\rho$ monotonically increases $(w<-1)$ in the first stage, and thereafter monotonically decreases $(w>-1)$, known as the "Quasi rip" \cite{Wei:2012ct}. At first, it thus tends to disintegrate bound structures in the universe, but then in the second stage the disintegration becomes reversed, implying that already disintegrated structures have the possibility to be recombined again. As an example model for this, we consider the energy density of dark energy to be a function of the scale factor and consider it's anastz to be \begin{equation}
\rho(a) = \rho_{0} a^{\alpha - \beta \ln{a} }
\end{equation} where a is the scale factor, $\alpha$ and $\beta$ are constants with $\rho_{0}$ being the energy density at some past time $t_{0}$.  Yet another possibility is the Little sibling of the big rip \cite{Bouhmadi-Lopez:2014cca} wherein the Hubble rate and the scale factor blow up but the  derivatives of the Hubble rate does not. This takes place at an infinite cosmic time with the  scalar curvature blowing up too. An example model for this also involves taking the energy density of dark energy as a function of the scale factor, given by \cite{Bouhmadi-Lopez:2014cca} \begin{equation}
\rho (a) = \Lambda + A \ln{\frac{a}{a_{0}}}
\end{equation}
\begin{table}[ht]
	\centering
	\caption{Comparison of Rip Scenarios and Example Models}
	\begin{tabular}{p{4.5cm} p{6.5cm} p{6.5cm}}
		\toprule
		Scenario & Description & Example Model \\
		\midrule
		Little Rip (LR) & Gradual energy density growth ($\rho$) over infinite time, asymptotically approaching a singularity. & $H(t) = H_{0} \exp{\lambda t}$ \\
		& & $H(t) = H_{0} \exp{C \exp{\lambda t}}$ \\
		\midrule
		Pseudo-Rip & Expansion accelerates with $H$ approaching a constant ($H_\infty$) but finite value. & $H(t) = H_{0} - H_{1} \exp{-\lambda t}$ \\
		\midrule
		Quasi Rip & Dark energy density $\rho$ first increases $(w < -1)$ and then decreases $(w > -1)$, implying disintegration and recombination of structures. & $\rho(a) = \rho_{0} a^{\alpha - \beta \ln{a}}$ \\
		\midrule
		Little Sibling of the Big Rip & Hubble rate and scale factor diverge, but derivatives of Hubble rate do not, with scalar curvature divergence. & $\rho(a) = \Lambda + A \ln{\frac{a}{a_{0}}}$ \\
		\bottomrule
	\end{tabular}
	\label{tab:rip_scenarios}
\end{table}
Table \ref{tab:rip_scenarios} summarizes all these various scenarios as discussed till now. Furthermore, there have been loads of works which have explored all these alternative rip scenarios in non-standard cosmologies similar to how other singularities have also been probed in such models, the possibilites ranging from various modified gravity theories to holographic cosmologies and viscous models too \cite{Brevik:2011mm,Brevik:2021sgw,Brevik:2020nuo,Brevik:2015pda,Brevik:2014eya,Brevik:2012ka,timoshkin2023little,lohakare2022rip,BorislavovVasilev:2021srn,Pati:2021zew,BorislavovVasilev:2021elf,Astashenok:2012kb,Granda:2011kx,Brevik:2012nt,Houndjo:2012hj,Saez-Gomez:2012uwp,Albarran:2016ewi,Sarkar:2021izd,Balakin:2012ee,Matsumoto:2017gnx,Pati:2021zew,Meng:2012mb}. There have also been works which have discussed ways of avoiding or moderating these singularities \cite{Morais:2016bev,Belkacemi:2011zk,Bouhmadi-Lopez:2017kvc,Xi:2011uz} but we will not be going over the details of that here. 

\bibliography{references}
\bibliographystyle{unsrt}

\end{document}